\documentclass{aa}             

\usepackage{natbib}
\bibpunct{(}{)}{;}{a}{}{,} 

\input epsf

\usepackage{graphicx}
\usepackage{txfonts}

\begin{document}

\title{
Age and helium content of the open cluster NGC\,~6791\\
from multiple eclipsing binary members
\thanks{Based on observations carried out at the 
Nordic Optical Telescope at La Palma and ESO's VLT/UVES
ESO, Paranal, Chile (75.D-0206A, 77.D-0827A, 081.D-0091).}
\thanks{Tables 11--22 are only available in electronic form
at the CDS via anonymous ftp to cdsarc.u-strasbg.fr
(130.79.128.5) or via http://cdsweb.u-strasbg.fr/cgi-bin/qcat?J/A+A/}
} 
\subtitle{I. Measurements, methods, and first results}

\author{
K. Brogaard       \inst{1}
\and H. Bruntt    \inst{1} 
\and F. Grundahl  \inst{1} 
\and J. V. Clausen \inst{2}
\and S. Frandsen  \inst{1}
\and D. A. VandenBerg	\inst{3}
\and L. R. Bedin	\inst{4}
}

\offprints{K. Brogaard, e-mail: kfb@phys.au.dk}

\institute{
Department of Physics and Astronomy,
Aarhus University, 
Ny Munkegade, DK-8000 Aarhus C, Denmark
\and
Niels Bohr Institute, Copenhagen University,
Juliane Maries Vej 30, DK-2100 Copenhagen {\O}, Denmark
\and
Department of Physics and Astronomy,
University of Victoria,
P.O. Box 3055, Victoria, B.C., V8W 3P6, Canada
\and
Space Telescope Science Institute, 
3700 San Martin Drive, Baltimore, MD 21218, USA
}

\date{Received xx XX 2010 / Accepted xx XX  2010, version 27.09.2010}
 
\titlerunning{Eclipsing binaries in the open cluster NGC\,6791}
\authorrunning{K. Brogaard et al.}

\abstract
{Models of stellar structure and evolution can be constrained by measuring accurate parameters of
 detached eclipsing binaries in open clusters. Multiple binary stars provide the means to determine helium abundances in these old stellar systems, and in turn, to improve age estimates.
}
{Earlier measurements of the masses and radii of the detached eclipsing binary V20
in the open cluster NGC~6791 were accurate enough to demonstrate that there are significant differences between current stellar models. Here we improve on those results and add measurements of two additional detached eclipsing binaries, the cluster members V18 and V80. The enlarged sample sets much tighter constraints on the properties of stellar models than has hitherto been possible, thereby improving both the accuracy and precision of the cluster age.
}
{We employed (i) high-resolution UVES spectroscopy of V18, V20 and V80 to determine their spectroscopic effective temperatures, [Fe/H] values, and spectroscopic orbital elements, and (ii) time-series photometry from the Nordic Optical Telescope to obtain the photometric elements.}
{The masses and radii of the V18 and V20 components are found to high accuracy, with errors on the masses in the range $0.27$--$0.36$\% and errors on the radii in the range $0.61$--$0.92$\%. V80 is found to be magnetically active, and more observations are needed to determine its parameters accurately. The metallicity of NGC~6791 is measured from disentangled spectra of the binaries and a few single stars to be [Fe/H]$ = +0.29 \pm 0.03$ (random) $\pm$\ 0.07 (systematic). The cluster reddening and apparent distance modulus are found to be $E(B-V) = 0.160\pm0.025$ and $(m-M)_V = 13.51 \pm0.06$ . 
A first model comparison shows that we can constrain the helium content of the NGC~6791 stars, and thus reach a more accurate age than previously possible. It may be possible to constrain additional parameters, in particular the C, N, and O abundances. This will be investigated in paper II. 
}
{Using multiple, detached eclipsing binaries for determining stellar cluster ages, it is now possible to constrain parameters of stellar models, notably the helium content, which were previously out of reach. By observing 
 a suitable number of detached eclipsing binaries in several open clusters, 
 it will be possible to calibrate the age--scale and the helium enrichment parameter $\Delta\,Y/\Delta\,Z$, and provide firm 
 constraints that stellar models must reproduce.
}
\keywords{
Open clusters: individual \object{NGC 6791} --
Stars: evolution --
Stars: binaries: spectroscopic --
Stars: binaries: eclipsing --
Techniques: spectroscopy --
Techniques: photometry
}
\maketitle

\section{Introduction}
\label{sec:intro}

The open cluster NGC~6791 is interesting for several reasons. The main scientific motivation for studying this cluster is that it is one of the oldest and, at the same time, the most metal-rich open clusters known \citep{Origlia06,Carretta07,attm07}, and thus very important for understanding chemical evolution. In addition, it is a particularly populated open cluster, with stars in all stages of evolution from the main sequence to the white dwarfs \citep{King05, Bedin05, Bedin08}, as well as with numerous variable stars \citep{Bruntt03,Mochejska02,deMarchi07}. 

Despite a number of studies that have presented well-calibrated photometry and high-precision colour-magnitude diagrams (CMDs) (\citealt{King05,SBG03}, hereafter SBG03), the age of NGC~6791 has remained very uncertain until recently because of correlated uncertainties in distance, reddening, colour-temperature transformations, and metallicity.

 It is widely appreciated that detached eclipsing binaries 
 offer the possibility of determining accurate (and precise) masses and radii for the 
 system components, nearly independent of model assumptions \citep{Andersen91,Torres10}. If the binary resides in a star cluster, and one or both of its components are close to the turn-off, it is possible to put tight constraints on the age of the system by comparing the position of the primary and secondary in a mass--radius (MR) diagram to 
 theoretical isochrones. For stellar clusters, such an analysis has some
 significant advantages: the determination of the masses and radii is 
 independent of the usual uncertainties such as reddening and distance. Furthermore, since the 
 comparison to models is carried out in the MR diagram, one avoids
 the difficult process of transforming the effective temperatures and 
 luminosities of the models to observed colours and magnitudes. Thus, determining cluster ages in the MR diagram allows a direct confrontation between 
 observations and theory.
 
\citeauthor{Grundahl08} (2008; hereafter GCH08) showed that using this method with their measurements of the cluster member eclipsing binary V20, they could determine a precise cluster age with an error of only $\pm0.3$ Gyr for a given stellar model. However, they were unable to determine which of the models (if any) to trust, because the difference in predicted age due to the specific stellar model adopted was about four times greater than their measurement precision. 

 Here we undertake an analysis of three detached eclipsing 
 binary systems, V18, V20 and V80, in NGC\,6791, and determine accurate masses and radii for the components of two of the systems. We also measure spectroscopic $T_{\rm eff}$\ and $[\mathrm{Fe/H}]$\ values 
 from disentangled spectra of the binary stars. This is used, together with the cluster CMD, to demonstrate how multiple cluster member eclipsing binary measurements constrain stellar models and cluster parameters, like age and helium content, better than previously possible.

\section{Targets}
\label{sec:targets}

\begin{figure}
\epsfxsize=95mm
\epsfbox{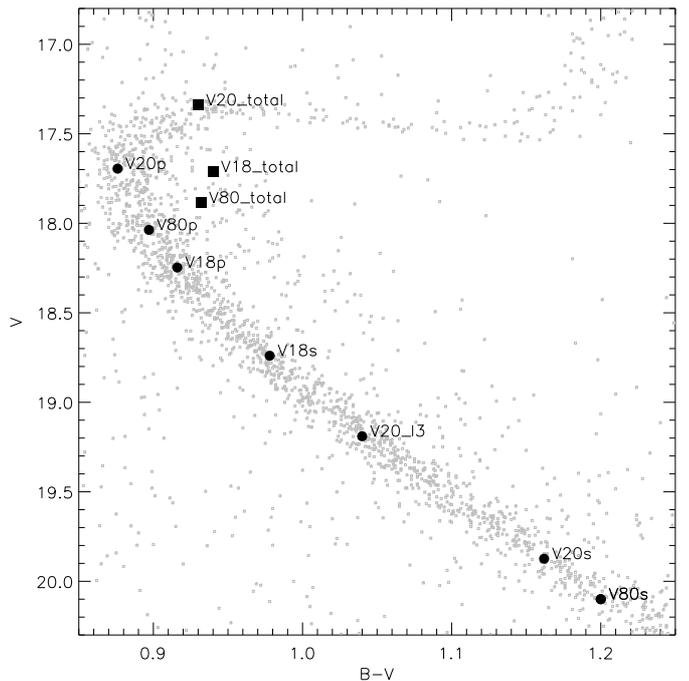}
\caption[]{\label{fig:dEBCMD}
The eclipsing binaries V18, V20 and V80 in the CMD of NCG~6791. Squares are the total system light, for V20 including a third light ($l_3$). Circles represent the calculated positions of the individual components as determined in this paper from light curve analysis and the assumption that all components are positioned on the cluster main sequence. $p, s,$ and $l_3$ indicate primary, secondary, and third light components, respectively.
}
\end{figure}

The detached eclipsing binaries analysed here are V18, V20 and V80, with names, alternative identifications, and coordinates given in Table \ref{tab:positions}, along with finding charts in appendix A. These systems were first identified as eclipsing binaries by \cite{Rucinski96} (V18 and V20) and \cite{Bruntt03} (V80). Their observed locations in the $(B-V),V$ CMD are shown in Fig.~\ref{fig:dEBCMD} together with calculated locations of the individual components as determined in Sect.~\ref{sec:std_phot}.

\begin{table}   
\caption[]{\label{tab:positions}
Names and coordinates for the eclipsing binaries.}
\begin{center}    
\begin{tabular}{llcc} \hline   
\hline\noalign{\smallskip}    
Name   & Alternative name & RA(2000.0) & DEC(2000.0) \\ 
\noalign{\smallskip}
\hline
\noalign{\smallskip}
V18 & V565\,Lyr & $19:20:49.38$ & $+37:46:09.3$ \\
V20 & V568\,Lyr & $19:20:54.30$ & $+37:45:34.7$ \\
V80 & 	--    & $19:21:06.48$ & $+37:47:27.8$ \\
\noalign{\smallskip}  
\hline
\end{tabular}            
\end{center}            
\end{table}    

\section{Photometry}
\label{sec:phot}

The photometric data for \object{V18} and \object{V80} consists of $V$ 
(Johnson) and $R$ (Cousins) CCD observations from the 2.56m Nordic Optical 
Telescope (NOT) and its ALFOSC instrument. In order to reduce the readout
time, while still including both \object{V18} and \object{V80} in each frame, we 
limited the readout section of the CCD to half its full size, thus covering
a field--of--view of $6\arcmin28\arcsec$\,$\times$\,$3\arcmin12\arcsec$. We refer to the
telescope homepage for further 
information about the ALFOSC instrument\footnote{{\scriptsize\tt http://www.not.iac.es}}. 

Observations were carried out on 9 nights between May and August 2009. Most observations were obtained on nights when both systems showed eclipses, in order to make efficient use of telescope time, resulting in
a total of 580 and 538 exposures in $V$ and $R$, respectively. 
For all observations we employed an exposure time of 240s in $V$ and 180s 
in $R$.  The photometric data for \object{V20} used in this paper are the 
same as those by GCH08. That data set also contained an egress of the primary eclipse of \object{V18}, which was 
missed during the 2009 campaign, thereby allowing us to 
obtain complete coverage of both eclipses and better out--of--eclipse coverage 
in the $V$ light curve for this object. 

The bias frames and flat fields, used in the data reduction, were obtained 
during evening twilight on each observing night.  All photometry was carried 
out with DAOPHOT/ALLSTAR/ALLFRAME \citep{stetson87, stetson94} and 
transformed to a common coordinate system using MATCH and MASTER (P. Stetson, 
private comm.). Each frame was processed using a point-spread function (PSF) 
calculated from about 200 of the brightest stars in the field.

\subsection{Light curves and standard indices}
\label{sec:lc}

With the instrumental photometry in hand, we proceeded to transform the 
observations to the $V$ and $R$ standard system using the same methodology 
as in GCH08. SBG03 put a large effort into the transformation of $BVI$ magnitudes onto the standard system, and 
we have used the available $V$ photometry from this source as internal 
standard stars. For each frame, a linear transformation from instrumental 
magnitudes to standard magnitudes was calculated, using $(B-I)$ (available from SBG03 for all stars in the field) as the colour term. Subsequently we averaged the coefficient for the colour term for all frames (for a given filter) and used this for the final 
determination of the zeropoint for each frame. The $BVI$ photometry listed in Table~\ref{tab:photometry}
is therefore on the same system as SBG03.
The accuracy of the photometric zeropoint is in the range 0\fm01 to 0\fm02 
as mentioned in SBG03. Our source for $R$ standard magnitudes was the photometry 
by \cite{Mochejska05}, as downloaded from the PISCES 
homepage\footnote{{\scriptsize\tt http://users.camk.edu.pl/mochejsk/PISCES/data.html}}. Apart from the different source of standard magnitudes, the procedure followed was identical to that for $V$. \cite{Mochejska05} reports an offset in their $V$ photometry of $0.047$ relative to SBG03, and therefore the $(V-R)$ colours in Table~\ref{tab:photometry} may not be very accurate.

\begin{figure}
\epsfxsize=90mm
\epsfbox{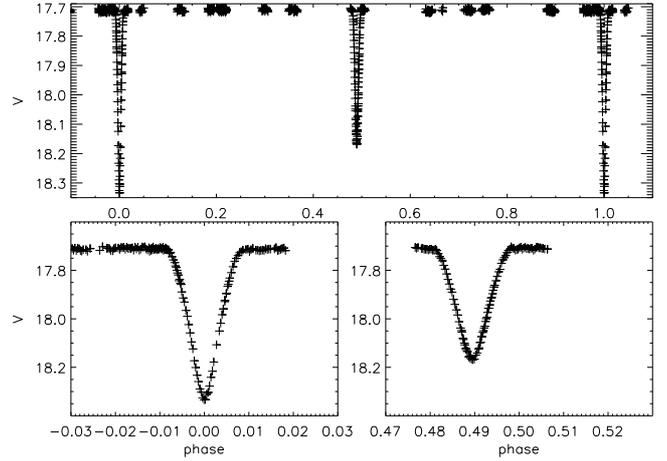}
\caption[]{\label{fig:v18_V}
Phased $V$ light curve for V18. The secondary eclipse is shifted away from phase 0.5, showing that the orbit is eccentric. 
}
\end{figure}

\begin{figure}
\epsfxsize=90mm
\epsfbox{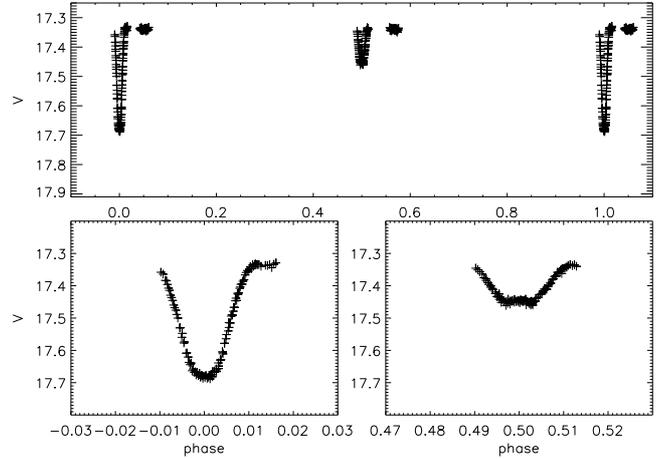}
\caption[]{\label{fig:v20_V}
Phased $V$ light curve for V20 from GCH08. The secondary eclipse is at phase 0.5 implying that the orbit is circular. The secondary eclipse is total. 
}
\end{figure}

The $V$ and $R$ light curves of V18 and V80 are listed in Tables~11-14 on CDS. A few obvious outliers and systematically deviating measurements at dusk and dawn have been removed. $V$ light curves of V18, V20 and V80 are shown in Fig.~\ref{fig:v18_V}--\ref{fig:v80_V} with phases calculated from the ephemerides given in Eqs.~\ref{eq:v18_eph}--\ref{eq:v80_eph} (Sect.~\ref{sec:eph}). For V20 we reuse the light curves from GCH08. In the top panels, all systems are shown on the same scale for easy comparison.

Throughout the paper, the component eclipsed at the deeper eclipse at 
phase 0.0 is referred to as the primary $(p)$, and the other as the 
secondary $(s)$ component.

V18 and V20 are seen to be well detached with a practically constant light level 
outside eclipses, while V80 shows signs of magnetic activity, most likely due to its shorter orbital period. For V20 
and V80 the secondary eclipse occurs at phase 0.50, and the eclipses are of equal duration, suggesting that their orbits are circular. V18 has a slightly eccentric orbit, as seen by the offset from phase 0.5 of the secondary eclipse.

\subsection{A new measurement of the third light component of V20}
\label{sec:v20_third}

For V20, a very close companion is included in the light curves, 
meaning that a significant amount of third light is present. 
GCH08 estimated the amount of third light by 
finding a solution where the total light of the system is shared 
among three stars constrained to lie on the cluster main sequence. However, they also 
mention that high-resolution imaging would allow a direct determination 
of the contribution to the total light from the third star. We therefore identified V20 in ACS images from the Hubble Space Telescope ({\it HST}), measured the magnitudes of the light
   emitted by the third star, which is well separated from the binary,
   and transformed the resultant magnitudes in the {\it HST} filters to $V$
   and $I$. To accomplish the latter, we selected stars bearing the
   closest similarity to the third light that we could find; specifically,
   stars within 0.003 mag in the ($m606 - m814$) colour index and within
   0.05 mag in $m606$.  The four twins to the third light that were
   identified this way, in our {\it HST} photometry, were located in the ground-based
   observations of SBG03, from which their $B$, $V$, and $I$ magnitudes
   were determined.

 The mean and rms spread of these four stars is our measurement of the third 
light magnitudes in these bands. We find $B=20.23\pm0.01$, $V=19.19\pm0.01$ 
and $I=18.12\pm0.01$.  Since we only have four stars these error estimates 
could be underestimated and we therefore conservatively adopt an error 
of $\pm0.02$ for all bands.
From these and the magnitudes of the combined light of V20 we calculated 
the third light contribution to V20, $l_{3}=\frac{L_{3}}{L_{p}+L_{s}+L_{3}}$, which is 
$0.164\pm0.004$ in $B$, $0.182\pm0.004$ in 
$V$ and $0.197\pm0.004$ in $I$. These values are significantly higher and much more accurate than
$0.146\pm0.022$ in $V$ and $0.168\pm0.025$ in $I$ as found indirectly by GCH08.

\subsection{Standard photometry of the binary components}
\label{sec:std_phot}
Standard, out--of--eclipse, $B,V,R,I$ photometry for the binaries is listed in 
Table~\ref{tab:photometry} together with calculated individual 
photometry for the stars from their light curve solutions and, in addition for $B-V$, the assumption that all stars of each system are located on the cluster CMD fiducial (as done for V20 in GCH08). For V80 the colours are calculated using only this assumption, which results in $L_{s}/L_{p}=0.15\pm0.05$ in $V$, since the light curve does not constrain the light ratio of this system. 


\begin{figure}
\epsfxsize=90mm
\epsfbox{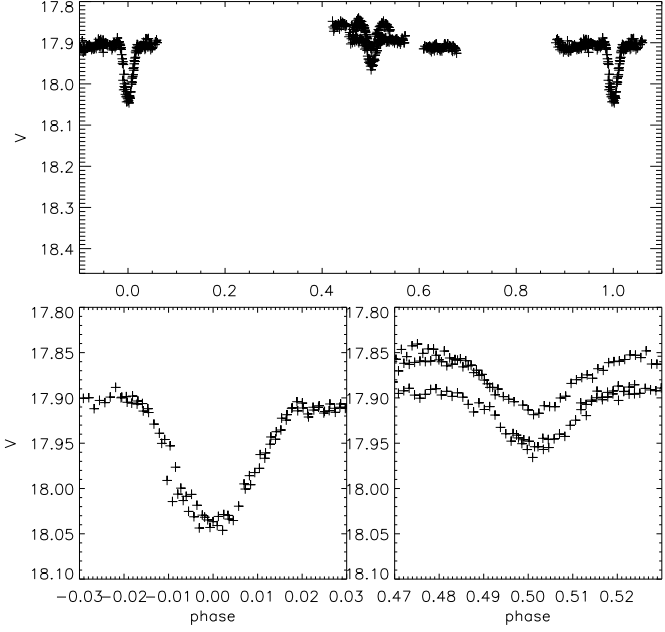}
\caption[]{\label{fig:v80_V}
Phased $V$ light curve for V80. The secondary eclipse is at phase 0.5, implying that the orbit is circular. Variations in the overall magnitude level indicate magnetic activity. 
}
\end{figure}


\begin{table}   
\caption[]{\label{tab:photometry}
Standard photometry for the eclipsing binaries and their components.
}
\begin{center}    
\begin{tabular}{lrrrr} \hline   
\hline\noalign{\smallskip}    
Name   & \multicolumn{1}{c}{$V$} & \multicolumn{1}{c}{$B-V$} & \multicolumn{1}{c}{$V-R$} & \multicolumn{1}{c}{$V-I$}\\ 
\noalign{\smallskip}
\hline
\noalign{\smallskip}
$V18_{total}$		&	17.713	&	0.940	&	0.505	&	0.990	\\
$V18_{primary}$		&	18.247	&	0.916	&	0.478	&	(...)	\\
$V18_{secondary}$	&	18.740	&	0.978	&	0.546	&	(...)	\\
\noalign{\smallskip}    
$V20_{total}$		&	17.340	&	0.930	&	(...)	&	0.985	\\
$V20_{primary}$		&	17.695  &	0.876	&	(...)	&	0.923	\\
$V20_{secondary}$	&	19.874	&	1.162	&	(...)	&	1.246	\\
$V20_{l_3}$		&	19.190  &	1.040	&	(...)	&	1.070	\\
\noalign{\smallskip}
$V80_{total}$		&	17.886	&	0.932	&	(...)	&	1.015	\\
$V80_{primary}$		&	18.037	&	0.897	&	(...)	&	(...)	\\
$V80_{secondary}$	&	20.100	&	1.200	&	(...)	&	(...)	\\

\noalign{\smallskip}  
\hline
\end{tabular}            
\end{center}
\tablefoot{
The reason some {\bf colour} indices are missing is that the we do not have light curves and light ratios for all systems in all colors. For $(B-V)$ we have calculated the {\bf colour} indices by finding solutions where all components are constrained to the main sequence. This is done to be able to put all systems on the same CMD (Fig.~\ref{fig:dEBCMD}).            
}
\end{table}                                  
\subsection{Periods and ephemerides}
\label{sec:eph}

For V20 we adopted the ephemeris of GCH08. For V18 and V80 we produced phased diagrams adding small shifts to the spectroscopic periods in order to determine the best initial estimates for the periods and ephemerides. 
We used these initial estimates to determine both the epoch and the period from analysis of our $V$ and $R$ light curves (see Sect.~\ref{sec:phel}). 
For each system the periods obtained from the two light curves agree well and are very close to the initial estimates.

We adopt the following linear ephemerides for all analysis in this paper: 

\begin{equation}
\label{eq:v18_eph}
\begin{tabular}{r r c r r}
{\rm Min \,I (V18)} =  & 2454651.4506  & + & $18\fd 798638$ &$\times\; E$ \\
                    &       $\pm 1$ &   &        $\pm10$ &             \\
\end{tabular}
\end{equation}

\begin{equation}
\label{eq:v20_eph}
\begin{tabular}{r r c r r}
{\rm Min \,I (V20)} =  & 2453151.6061  & + & $14\fd 469918$ &$\times\; E$ \\
                    &       $\pm 9$  &  &        $\pm25$ &             \\
\end{tabular}
\end{equation}

\begin{equation}
\label{eq:v80_eph}
\begin{tabular}{r r c r r}
{\rm Min \,I (V80)} =  & 2454652.3045  & + & $4\fd 88594$ &$\times\; E$ \\
                    &      $\pm 13$ &   &      $\pm16$ &             \\
\end{tabular}
\end{equation}

\subsection{Photometric elements}
\label{sec:phel}
V18 and V20 are well detached systems with no signs of out--of--eclipse variability. Furthermore, the components of each system are small relative to their separation. Thus, they are well suited for accurate measurements. The light curves of V80 show signs of magnetic activity (Fig.~\ref{fig:v80_V}), and more observations are needed before we can hope to obtain precise and accurate photometric elements for this system. However, since we can already get a good measurement of $\log g$, which is useful for the further spectroscopic study, we include preliminary dimensions of this system as well. 

We adopted the simple Nelson-Davis-Etzel model \citep{Nelson72,Etzel81} for the light curve analysis. 
It represents the deformed stars as biaxial ellipsoids and applies a simple 
bolometric reflection model.
We have used the corresponding 
JKTEBOP\footnote{{\scriptsize\tt http://www.astro.keele.ac.uk/$\sim$jkt/}} code, which is a
revised and extended version of the original EBOP code \citep{Etzel81}. 
The Levenberg-Marquardt minimization algorithm (MRQMIN: \citealt{Press92})
is used for the least-squares optimization of the parameters, and the code has
been extended to include non-linear limb darkening and adjustment of
epoch and orbital period. We made use of important new features in the latest version, namely the possibility to fit the light curve simultaneously with an externally determined light ratio \citep{Southworth07}, a third light ratio \citep{Southworth10}, and $e\sin\omega$ and $e\cos\omega$ \citep{Southworth09}, including their errors. 
In some of its modes, JKTEBOP performs Monte Carlo simulations \citep{Southworth04b,Southworth04} and residual shift analysis \citep{Southworth08}, which we used to assign realistic errors to the photometric elements.

The magnitude at quadrature was always included
as a free parameter, and the phase of primary eclipse was allowed to
shift from 0.0. In initial JKTEBOP analysis, the epoch and orbital period
were included as free parameters and then fixed at the values given
in Eqs.~\ref{eq:v18_eph}--\ref{eq:v80_eph}. A circular orbit
was assumed for V20 and V80 as found from the spectroscopic orbits. For V20, a light curve analysis with the eccentricity as a free parameter also resulted in a circular orbit supporting the assumption, while the V80 light curve is not of a high enough quality to make such a test. The mass ratios, $q$, between the components were kept at the
spectroscopic values (Table~\ref{tab:orbit}).  
Gravity darkening coefficients corresponding to
convective atmospheres were applied, and the simple bolometric reflection
model built into JKTEBOP was used. Tests showed that these assumptions have negligible effect on the derived photometric elements.

The procedure we followed in order to minimize the errors related to limb darkening deserves some attention. It is well known that the linear limb-darkening law is a poor fit to both the observed limb darkening of the Sun and that predicted by theoretical model atmospheres. We therefore adopted a non-linear limb-darkening law, more precisely the square root law, since according to van Hamme (1993) this gives the best fit to stellar atmospheres in the temperature range of these binary stars. Theoretical coefficients to be used with the square root law for a given star have been calculated by several authors using 1D stellar atmospheres (e.g. \citealt{vh93,Claret00}). They all depend on the $T_{\rm eff}$, $\log g$, and $[\mathrm{Fe/H}]$. Furthermore, they depend on the stellar atmosphere model used for the calculation. To make things more complicated it has been shown that 1D stellar atmospheres do not reproduce the limb darkening of the Sun (\citealt{Sing08}), and therefore theoretical limb-darkening coefficients will be inaccurate in any case. In order to minimize the errors related to these facts, the optimal approach would be to fit the coefficients as part of the light curve solutions, instead of fixing them at theoretical values. But most light curves, including ours, are not of sufficient quality to fit for both a linear and non-linear coefficient for both stars.   
Instead we exploit a fact demonstrated by \cite{Southworth07} that for the square root law, the linear and non-linear coefficients are highly correlated. Therefore, by fixing one coefficient and fitting for the other, any error in the fixed coefficient will be compensated for by a shift in the value of the fitted coefficient. This is the approach we adopted, since it allowed us to use a non-linear law while minimizing dependence on the model and the exact stellar parameters. 

In the text and tables with photometric solutions we use the following symbols:
$i$ orbital inclination;
$r_p = R_p/a$; relative radius of primary;
$r_s$ relative radius of secondary;
$k =  r_s/r_p$;
$J$ central surface brightness ratio;
$L$ luminosity;
$l_3$ third light fraction;
$e$ eccentricity;
$\omega$ longitude of periastron.

\subsubsection{V18 photometric elements}
\label{sec:V18phel}
Our first photometric solutions for V18 showed that $k$ is not well constrained by the light curve without additional external constraints. This is typical for a system with a slightly eccentric orbit without a total eclipse. We therefore found solutions for the $V$ band light curve which simultaneously fit the light curve and spectroscopic values of $e\sin\omega$ and the light ratio (Sect.~\ref{sec:rv} and \ref{sec:specl}), including their errors. We could not repeat this procedure for the $R$ band, since we have no spectroscopic measurement of the light ratio corresponding to this band. Therefore we adopt the $V$ band solution for our final measurements. To get the $R$ band surface-brightness ratio and luminosity ratio, we fit the $R$ band light curve with the other parameters fixed from the $V$ band solution. Table \ref{tab:v18_ebop} shows our adopted photometric elements. Errors are determined using JKTEBOP Monte Carlo simulations, since residual shift errors indicated that correlated errors are not present at a significant level in the observed light curves. Square root limb-darkening coefficients from \cite{Claret00} for our spectroscopically measured metallicity and effective temperatures were used as a starting point. We followed the previously mentioned strategy of fixing either the linear or non-linear coefficient and solving for the other. The resulting elements remained identical to well below 0.01\% whether we solved for the linear or non-linear coefficient. Further tests showed that fixing both the linear and non-linear limb-darkening coefficients instead of employing the spectroscopic $e\sin\omega$ constraint changes the relative radii by only 0.13\% and 0.04\% for the primary and secondary component respectively. This seems to suggest that at least in this case the theoretical limb-darkening coefficients are reliable. 
As seen in Table \ref{tab:v18_ebop}, the relative radii of V18 have been measured with errors of 0.6\% and 0.9\% for the primary and secondary components respectively.

\begin{table}
\caption[]{\label{tab:v18_ebop}
Photometric solution for V18 from the JKTEBOP code.
}
\begin{center}
\begin{tabular}{lr} \hline
\hline\noalign{\smallskip}
Parameter            &	Value\\ 
\hline\noalign{\smallskip}        
Spectroscopic constraints & \\
\noalign{\smallskip}                 
$L_s/L_p$ ($V$)      & $0.636\pm0.020$    \\
\noalign{\smallskip}                 
$e\sin\omega$	     & $0.0100\pm0.0011$  \\
\hline\noalign{\smallskip}                        
Measured parameters & \\
\noalign{\smallskip}
$i$ \,(\degr)        & $89.377\pm0.022 $\\
\noalign{\smallskip}                 

$r_p + r_s$          & $0.0560\pm0.0002$\\
\noalign{\smallskip}
                     
$k = r_s/r_p$        & $0.882\pm0.011$\\
\noalign{\smallskip}

$r_p$                & $0.02975\pm0.00018$\\
\noalign{\smallskip}

$r_s$                & $0.02623\pm0.00024$\\
\noalign{\smallskip}

$e$		     & $0.0193\pm0.0006$ \\
\noalign{\smallskip}

$\omega$	     & $149.1\pm 2.8$ \\
\noalign{\smallskip}

$J_s/J_p (V)$        & $0.817\pm0.018$\\
\noalign{\smallskip}
               
$J_s/J_p (R)$        & $0.791\pm0.074$\\
\noalign{\smallskip}
                     
$L_s/L_p (V)$        & $0.635\pm0.016$    \\
\noalign{\smallskip}
$L_s/L_p (R)$        & $0.674\pm0.042$    \\
\noalign{\smallskip}
           
$\sigma$ \, ($V$-mmag.)   &  4.36  \\
\noalign{\smallskip}            
\hline
\end{tabular}            
\end{center}
\end{table}                       
\subsubsection{V20 photometric elements}
\label{sec:V20phel}

V20 was analysed by GCH08. However, we measured more directly and accurately the third light and found it to be outside their error estimate. We therefore performed a new analysis using their light curve. Table~\ref{tab:v20_ebop} shows our measured photometric elements for each band and our adopted solution. 
From the JKTEBOP Monte Carlo and residual shift solutions we found that significant correlated errors are present in the observed light curves, and we chose to adopt the larger residual shift errors as our error estimates. For the final parameters we adopted an error weighted mean of the light curve solutions from each band. Due to the systematics present, we do not allow the error of the adopted solution to be reduced by averaging the solutions from the two bands, but adopt the error from the $V$ band as our final estimate.   
As seen in Table~\ref{tab:v20_ebop}, relative radii of V20 have been measured with errors of 0.9\% and 0.7\% for the primary and secondary components respectively. Our measurement of $r_s=0.02519\pm0.00017$ is two sigma larger than $r_s=0.0248\pm0.0002$ found by GCH08. Although we also use a different limb darkening law than GCH08, this has minor effects on the solution. Our new measurement of the third light is the main reason we find a larger radius for the secondary. 

\begin{table}
\caption[]{\label{tab:v20_ebop}
Photometric solutions for V20 from the JKTEBOP code.
}
\begin{center}
\begin{tabular}{lrrr} \hline
\hline\noalign{\smallskip}
Parameter                 &  $V$       &      $I$         &	adopted\\ 
\hline\noalign{\smallskip}        
$i$ \,(\degr)        &  89.99 	&  89.99    & 89.99\vspace{-0.8mm}\\
                     & $_{-0.15}^{+0.01}$ & $_{-0.18}^{+0.01}$  & $_{-0.15}^{+0.01}$ \\
\noalign{\smallskip}
$r_p + r_s$          &  $0.0700$ &  $0.0710$   & $0.0702$\\
                     &  $\pm0.0005$ &  $\pm0.0010$  & $\pm0.0005$  \\
\noalign{\smallskip}
$k$		     &  0.559   &  0.559    & 0.560\vspace{-0.8mm}\\
                     &  $\pm0.005$ &  $\pm0.008$  & $\pm0.005$  \\

\noalign{\smallskip}
$r_p$                &  0.04491    & 0.04551  & 0.04501\\
                     &  $\pm0.00040$ &  $\pm0.00078$  & $\pm0.00040$  \\

\noalign{\smallskip}
$r_s$                &  0.02513    & 0.02546  & 0.02519\\
                     &  $\pm0.00017$ &  $\pm0.00026$  & $\pm0.00017$  \\

\noalign{\smallskip}
$J_s/J_p (V)$        &  0.414	& (...)	  & 0.410\\
                     &  $\pm0.043$ & (...)   & $\pm0.043$  \\

\noalign{\smallskip}
$J_s/J_p (I)$        & (...) 	&  0.591  & 0.590\\
                     & (...)    &  $\pm0.074$  & $\pm0.074$  \\

\noalign{\smallskip}
$L_s/L_p (V)$        &  0.134   &  (...)       & 0.134\\
                     &  $\pm0.003$ & (...)     & $\pm0.003$  \\

\noalign{\smallskip}
$L_s/L_p (I)$        & (...)         &  0.181  & 0.181\\
                     & (...)         &  $\pm0.005$  & $\pm0.005$  \\

\noalign{\smallskip}
$l_3 (V)$            &  0.182   &  (...)       & 0.182\\
\noalign{\smallskip}
$l_3 (I)$            & (...)         &  0.197  & 0.196\\
\noalign{\smallskip}
$\sigma$ \, (mmag.)  &  4.6     &  6.8    &   (...)\\

\noalign{\smallskip}            
\hline
\end{tabular}            
\end{center}
\end{table}                       
\subsubsection{V80 photometric elements}
\label{sec:V20phel}

As mentioned earlier, V80 shows signs of magnetic activity in the light curve and more observations are needed for very accurate measurements. In any case, to obtain a measurement of $\log g$\, to be used for $T_{\rm eff}$\ and $[\mathrm{Fe/H}]$\ measurements, we did a preliminary light curve analysis for V80. To be specific, we first rectified the light curves. In Fig.~\ref{fig:v80_V} both primary and secondary eclipses are covered by observations from two different nights. As seen the overall magnitude level remained identical for the nights covering the primary eclipse while it changed between the two nights covering the secondary eclipse. However, the depth of the secondary eclipse did not change between nights. The rectification we employed was therefore to shift the one secondary eclipse coverage which was offset relative to the rest, so that all eclipse observations were aligned in magnitude. Then deviating out--of--eclipse observations were also shifted to match this magnitude level. We then found JKTEBOP solutions employing the $V$ band light ratio determined from the CMD in Sect.~\ref{sec:std_phot}. The results for the relative radii and inclination are shown in Table \ref{tab:v80_ebop}. Errors are based on solutions of the $V$ band light curve varying the light ratio within the errors given by the external constraint and on the level of consistency between $V$ and $R$ band solutions. As expected, the errors are much higher than for the other systems, in the range of 6-20\%.

\begin{table}
\caption[]{\label{tab:v80_ebop}
Photometric solution for V80 from the JKTEBOP code.
}
\begin{center}
\begin{tabular}{lr} \hline
\hline\noalign{\smallskip}
Parameter            &	Value\\ 
\hline\noalign{\smallskip}        
Constraints from CMD & \\
\noalign{\smallskip}            
$L_s/L_p$ ($V$)      & $0.15\pm0.05$    \\
\hline\noalign{\smallskip}

Measured parameters  & \\
\noalign{\smallskip}            

$i$ \,(\degr)        & $84\pm1$\\
\noalign{\smallskip}            

$r_p$                & $0.0900\pm0.0054$\\
\noalign{\smallskip}            

$r_s$                & $0.061\pm0.012$\\
\noalign{\smallskip}            
\hline
\end{tabular}            
\end{center}
\end{table}                       
\section{Spectroscopy}
\label{sec:spec}

 The spectroscopic observations were carried out in service mode
 with UVES at the ESO VLT during allocation periods 75+77 (V20) and 81 (V18 and V80). Since NGC~6791
 is at declination $+37\degr$, it can be observed at Paranal only at an 
airmass higher than 2.1. Therefore all observations were carried out during the short observing window of a few hours
around meridian passage.
 Due to the faintness of the stars, and in order to minimize slit losses, a slit
 of 1\farcs20 width, corresponding to a resolution of approximately
 37\,000, was used. For the observations of V18 and V20, the slit was aligned along the parallactic angle
 (ELEV mode), and the atmospheric dispersion corrector (ADC) was not inserted in the beam, since it causes a
 slight loss of flux. The procedure for V80 was different, since a fixed slit position allowed us to
 put an additional star on the slit together with V80, to be used for a metallicity measurement. 
 The UVES standard 580nm setup, and on-chip binning of 2$\times$2 pixels,
was used for all observations. 
 The corresponding wavelength ranges covered by the two CCD detectors 
employed in UVES are approximately 4775--5750{\AA} and 5875--6830{\AA},
respectively. The typical S/N per pixel for the red chip was between 15 and 25.
A total of 13, 15, and 10 usable epochs were obtained for V18, V20 and V80, respectively, each followed by a ThAr exposure for wavelength calibration.

We used the UVES pipeline to reduce the spectra of all three binaries, with a few changes relative to the standard pipeline settings: We reduced the threshold to discard deviating pixels from 10 sigma to 3 sigma, and we used the semi-automatic flux- weighted wavelength calibration. For the observations with more than one star on the slit, we applied offsets and reduced the extraction slit length, to extract the spectrum of each star separately. We did not merge orders, since the regions of overlap have low S/N and could introduce unwanted noise in the later analysis.

V20 was previously analysed by GCH08 using the UVES pipeline reduced spectra as supplied by ESO. However, inspecting the raw spectra of V20 revealed that the choice of the ELEV mode had the unfortunate effect of allowing light from a close-by star (in addition to the third light) to reach the slit for some observations. This affects the radial velocity zero-point determination significantly, regardless of whether this is done using the third light component as in GCH08 or with our present method as we describe below. Therefore, these spectra were re-reduced with offsets and reduced slit length, to mask out the unwanted star. 

The orbital periods of V18 and V80 were not well known prior to the spectroscopic observations. Therefore, we retrieved the data from the ESO Archive the day after each new observation, reduced the spectrum, made an orbital solution to find a better period estimate, and then rescheduled the remaining observations to optimal epochs.   

\subsection{Radial velocities and spectroscopic elements}
\label{sec:rv}
For the radial velocity measurements we employed the broadening function (BF) formalism \citep{Rucinski99,Rucinski02,Rucinski04}. The BF method assumes that all components in the observed spectrum can be well described
by the same synthetic template spectrum with respect to the relative strengths of spectral lines. This assumption holds as long as the spectral types of the components do not differ too much. We used synthetic spectra from the grid of \cite{Coelho05} and repeated the procedure with templates covering a reasonable range in $\log g$, $T_{\rm eff}$\ and $[\mathrm{Fe/H}]$\ and found that the effect of the adopted template is much less than the final adopted error estimates. We also confirmed the similarity of the spectral types of the components by separating the spectra (see Sect.~\ref{sec:metal}).

\begin{figure}
\epsfxsize=90mm
\epsfbox{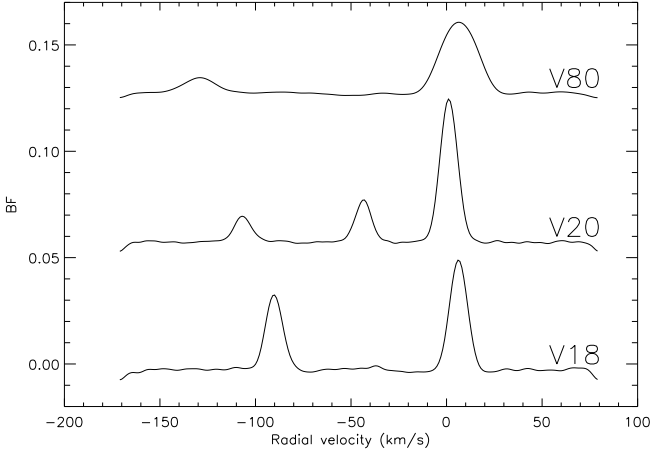}
\caption[]
{
\label{fig:bf}
Averaged broadening functions for the three binaries using the optimal templates and Gaussian smoothing, showing the characteristics of each system. V80, V20, and V18 BFs are shown from top to bottom with offsets of 0.13, 0.06, and 0.00 for clarity. V18 and V20 are seen to be slowly rotating systems, while V80 rotates faster. V20 contains a third light component.
}
\end{figure}

In our implementation BFs are calculated for each order and averaged to give the final BF profile. This profile is smoothed by convolution with a Gaussian. The width of this smoothing Gaussian was chosen to minimize the order--to--order scatter of radial velocities measured from single-order BFs. Fig.~\ref{fig:bf} shows example BFs for each system.

The final BFs were then fitted with a Gaussian around each peak to measure the radial velocities. We found that this procedure gives a lower order--to--order radial velocity scatter than fitting a rotational profile, when applied to the individual orders.
We did not use all spectral orders, since in the bluest region, the S/N was too low to obtain a decent BF, and some orders contain wide lines or telluric lines, and are therefore not well-suited for radial velocity measurements. We ended up using ten orders from each chip.

Spectroscopic elements were derived from the measured radial velocities 
using the method of Lehman-Filh\'es implemented in the SBOP program 
\citep{Etzel04}. The orbital periods were fixed at their photometrically determined values.

Our first solutions had $(O-C)$ errors of up to 1 km/s in a systematic pattern identical for the primary and secondary components. This indicated that the radial velocity errors were dominated by radial velocity zero-point offsets between epochs. Since our spectra are calibrated with ThAr spectra taken immediately after the stellar spectra, these offsets cannot be attributed to wavelength calibration errors. Instead they occur because the starlight is not exactly centred across the rather wide spectrograph slit, effectively causing an incident angle, which shifts the spectrum in wavelength relative to a star perfectly centred across the slit. To correct for this shift, we did the following: We identified a spectral order with very strong telluric absorption lines ($\lambda=6276-6299\AA$). In the adjacent orders we measured the BF profile, which is nearly identical for close-by orders. We then convolved our synthetic stellar template spectrum, covering the wavelength region with telluric absorption lines, with the obtained BF, to produce the expected stellar spectrum in that order. This spectrum was then multiplied with synthetic telluric spectra which were broadened to the spectrograph resolution, shifted in radial velocity from $-3$ to $+3$ km/s in steps of 50 m/s, and multiplied with factors between 0.5 and 1.5 in steps of 0.1. The resulting set of artificial spectra of stellar+telluric lines were all cross-correlated with the observed spectra, and the one giving the highest cross correlation determined the shift of the telluric lines. This shift is our radial velocity zero-point correction, since the telluric absorption lines are always at zero radial velocity (except for small shifts due to winds in the Earth's atmosphere), and the observed shift can only be caused by the aforementioned imperfect slit centring.

Applying the zero-point radial velocity shifts reduced the errors on the orbital parameters very significantly. It also changed the orbital parameters to just outside their original one-sigma error bars, indicating that these were underestimated because the zero-point shifts had not been taken into account.

In order to check that the errors on the spectroscopic elements given by the SBOP program for our zero-point corrected radial velocities can be trusted we developed a Monte Carlo routine for alternative error calculations. We assumed that the radial velocity errors have two Gaussian components, one of which represents any remaining radial velocity zero-point error and is identical for both components. We subtracted in quadrature the standard deviation radial velocity errors found from order--to--order scatter from the SBOP $(O-C)$ standard errors in order to estimate the magnitude of the radial velocity zero-point error. We found this error contribution to be $\sim$ 150 m/s. We then ran SBOP a large number of times adding to each radial velocity two Gaussian errors with standard deviations of 150 m/s and the order--to--order standard deviations, respectively. From the output solutions we calculated the one-sigma error. We found that this approach gives errors consistent with the errors from SBOP, and concluded that these errors can be trusted. Adopting a conservative approach we have anyway chosen to adopt the Monte Carlo errors, which are slightly (1-10\%) larger than the SBOP errors. For more details on the error estimates we refer to \cite{Brogaard10}. 

\begin{figure}
\epsfxsize=100mm 
\epsfbox{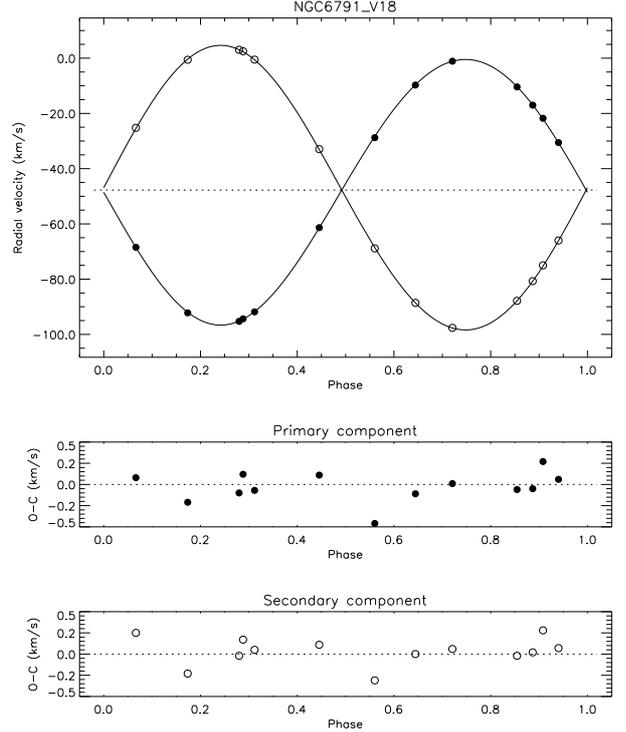}
\caption[]
{
\label{fig:sbopV18}
Spectroscopic double-lined orbital solution for V18. Phase 0.0 corresponds to central primary eclipse.
The horizontal dotted line (upper panel) represents the centre--of--mass velocity. }
\end{figure}

\begin{figure}
\epsfxsize=100mm 
\epsfbox{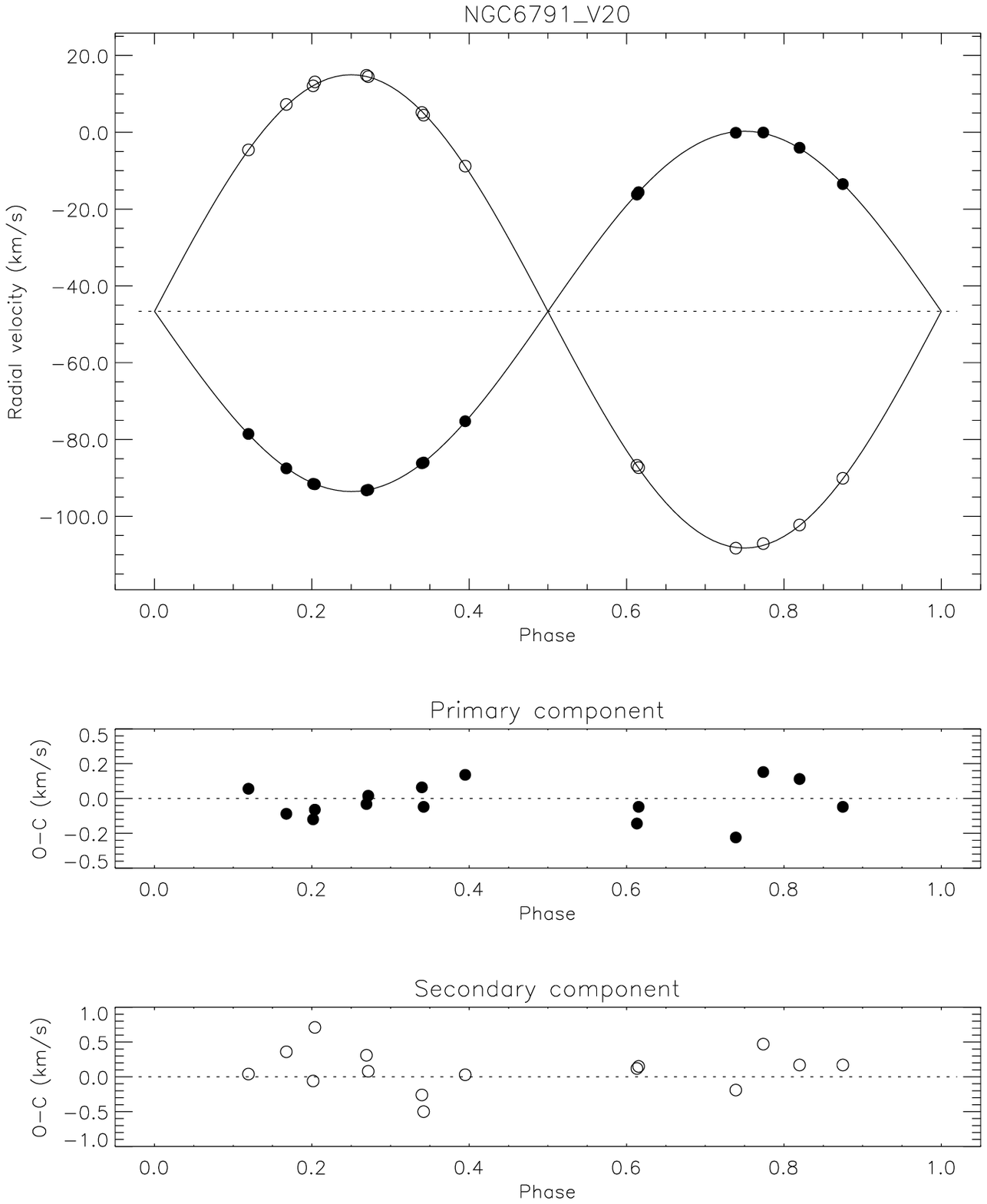}
\caption[]
{
\label{fig:sbopV20}
Spectroscopic double-lined orbital solution for V20. Phase 0.0 corresponds to central primary eclipse.
The horizontal dotted line (upper panel) represents the centre--of--mass velocity. Note the different scales in the $(O-C)$ panels.
}
\end{figure}

\begin{figure}

\epsfxsize=100mm 
\epsfbox{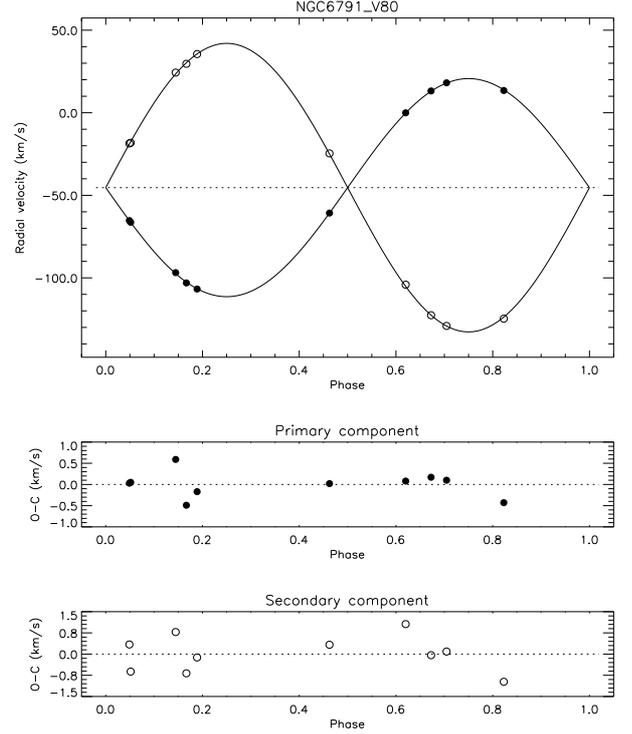}
\caption[]
{
\label{fig:sbopV80}
Spectroscopic double-lined orbital solution for the V80. Phase 0.0 corresponds to central primary eclipse.
The horizontal dotted line (upper panel) represents the centre--of--mass velocity. Note the different scales in the $(O-C)$ panels.
}
\end{figure}

We derived orbital parameters for the three binaries from both single- and double-lined solutions (meaning that a solution was found both by fitting one component at a time, and by fitting the two components in a combined solution). For the double-lined solutions, the radial velocities of the secondary were adjusted to account for the difference in gravitational redshift between components. These adjustments were calculated iteratively from the obtained absolute masses and radii and are $-35$ m/s for V18, $-180$ m/s for V20 and $-150$ m/s for V80. Furthermore, the relative radial velocity weights between components were set according to the difference in $(O-C)$ errors between components from single-lined solutions.
For V18 the orbit is slightly eccentric, so to obtain a consistent set of semi-amplitudes corresponding to the same $e$ and $\omega$ we adopted the double-lined solution. For this system, a measurement of $e\sin\omega$, including its error, was desirable for constraining the light curve solution, where this quantity is poorly determined. However, since $e$ and $\omega$ are correlated parameters, one cannot obtain the error on $e\sin\omega$ from the errors on $e$ and $\omega$. We therefore ran SBOP using Sterne's method, which fits for $e\sin\omega$ and $e\cos\omega$ directly. Sterne's method only works for one component, so we did a solution for both the primary and secondary component. This resulted in values for $e$ and $\omega$ on either side of the adopted solution. We therefore calculated the value of $e\sin\omega$ from the adopted $e$ and $\omega$ while we obtained the error on $e\sin\omega$ from the SBOP solutions using Sterne's method. This result is $e\sin\omega = 0.0100 \pm0.0011$.

For V20 we cannot rule out a low eccentricity of $\sim$ 0.002 from the spectra alone, but both $V$ and $I$ light curves prefer an orientation of 90 degrees when any eccentricity is forced upon them, indicating a strong preference for a circular orbit, which is what we adopt. Furthermore, we adopt the single-lined solutions since they correspond to a consistent set for a circular orbit, and this avoids any speculations about differences in observed system velocity between components due to exact differences in gravitational redshift, convective blueshifts and any template mismatch. For V18 such differences must be very small because the stars are of very similar masses, radii and spectral types, and therefore negligible errors are introduced by adopting the double-lined solution for this system.
For V80 rotational broadening reduces the accuracy of both radial velocities and zero-point measurements, but because of the higher radial velocity amplitudes caused by the shorter orbital period, the minimum masses are still determined to an accuracy better than 1\%.

We list the final spectroscopic elements for the systems in Table~\ref{tab:orbit} and show double-lined solutions in Figs.~\ref{fig:sbopV18}--\ref{fig:sbopV80}. Tables 15-17 containing radial velocities and ($O-C$) errors can be found on CDS. As seen, minimum masses accurate to well below 1\% have been obtained
for all systems. For V20, the errors on both the semi-amplitudes and individual measurements are about half of those found by GCH08. Perhaps the most convincing evidence, which indicate that our measurements are more accurate, is to compare our Fig.~\ref{fig:sbopV20} to their Fig. 8; note in particular the sets of measurements close to phases 0.2, 0.25, 0.35, and 0.6, which were taken immediately after each other and are therefore expected to give identical results. 

As also seen from Table~\ref{tab:orbit}, the systemic velocities, $\gamma$, of the systems are in excellent agreement
with the value of $-47.1 \pm 0.8$ km\thinspace s$^{-1}$\ as determined from 
15 cluster members by \cite{carraro06}, who also found the dispersion in the radial velocities of the 15 stars
to be $2.2 \pm 0.4$ km\thinspace s$^{-1}$.
This provides a strong argument that all systems are cluster members. V18 and V80 are in addition proper-motion members (96\% and 86\% probability according to Dr. Kyle Cudworth as stated by \citealt{Rucinski96} and \citealt{deMarchi07},  respectively). Given also the similarity in the metallicities that we derive (Sect.~\ref{sec:metal}) there should be no doubt that these stars are cluster members.  

\begin{table*}   
\caption[]{\label{tab:orbit}
Spectroscopic orbital solutions.
}
\centering
\begin{tabular}{lrrr} \hline   
\hline\noalign{\smallskip}    
Parameter & \multicolumn{1}{c}{V18} & \multicolumn{1}{c}{V20} & \multicolumn{1}{c}{V80} \\ 
\noalign{\smallskip}
\hline
\noalign{\smallskip}  
Fixed parameters:             & & &   \\
$T$~(HJD$-$2\,400\,000) &  54651.4506 &  53151.6061 &  54652.3045 \\
$P$~(days)              & 18.798638   & 14.469918   & 4.88594\\
\noalign{\smallskip}    
Free parameters:            & & &   \\ 
$K_p$~(km\thinspace s$^{-1}$)    &$ 48.094 \pm 0.075 $ &$ 46.900 \pm 0.050 $ &$ 66.05 \pm 0.16 $ \\ 
$K_s$~(km\thinspace s$^{-1}$)    &$ 51.523 \pm 0.076 $ &$ 61.59 \pm 0.10 $ &$ 87.38 \pm 0.26 $ \\
$\gamma$~(km\thinspace s$^{-1}$) &$-47.743 \pm 0.038 $ &$-46.639 \pm 0.035 $ &$-45.355 \pm 0.098 $ \\
$e$                     &  $0.0196 \pm 0.0012$	&  $0.00$(fixed)	&  $0.00$(fixed)   \\ 
$\omega$\,(\degr)	& $149.4 \pm 3.1$	& $90.00$(fixed)		& $90.00$(fixed)   \\	
\hline\noalign{\smallskip}  
Derived parameters:             &      \\

$M_p \sin^3i~\mathrm{(M_{\sun})}$       & $ 0.9953  \pm 0.0033  $ & $ 1.0868  \pm 0.0039  $ & $ 1.0415  \pm 0.0069  $\\
$M_s \sin^3i~\mathrm{(M_{\sun}})$       & $ 0.9291  \pm 0.0032  $ & $ 0.8276  \pm 0.0022  $ & $ 0.7872  \pm 0.0043  $\\
$q$					& $ 0.9334  \pm 0.0020  $ & $ 0.7615  \pm 0.0015  $ & $ 0.7559  \pm 0.0028  $\\
$a \sin i~\mathrm{(R_{\sun})}$          & $37.010  \pm 0.040   $ & $31.030 \pm 0.032  $ & $14.819   \pm 0.029  $\\

\noalign{\smallskip}  
Other quantities:              & & &    \\
$\sigma_p$(km\thinspace s$^{-1}$)	& 0.24	& 0.14	& 0.33   \\
$\sigma_s$(km\thinspace s$^{-1}$)	& 0.23	& 0.31	& 0.69   \\
$N_{\rm obs}$                      &    13 &    15 &    10\\
Time span (days)               &    88 &   377 &    92\\
\noalign{\smallskip}  
\hline
\end{tabular}            
\tablefoot{$T$ is the time of central primary eclipse.}          
\end{table*}                                  
\subsection{Spectroscopic light ratio of V18}
\label{sec:specl}

For V18 we found that a spectroscopic light ratio could improve the photometric elements (cf. Sect.~\ref{sec:V18phel}). For this system we therefore found an additional use for our calculated broadening functions, since the BF is a proxy for the mean line profiles in the stars. The difference in equivalent widths between each component peak therefore measures the light ratio. We selected BFs from 10 orders for which the mean luminosity ratio corresponds very closely to the luminosity ratio in the $V$ band according to tests with both Planck functions and synthetic spectra of Coelho et al. (2005). For each epoch of observations we fitted these BFs with a rotational profile convolved with a Gaussian and integrated the area under each peak to measure the ratio of equivalent widths between components. If the stars had identical spectral types this ratio measures directly the light ratio in the $V$ band. Since the light curve solutions suggest an effective temperature difference of about 200 K, a very small correction of $-0.06$ was applied to the measured light ratio. We found 
$L_{s}/L_{p}=0.636\pm0.020$ where the error is a sum of the epoch--to--epoch measurement scatter ($\pm0.005$), a contribution due to possible differences between light ratio in the $V$ band and the spectral region measured ($\pm0.005$), an estimated uncertainty in the correction due to different spectral types ($\pm0.005$), and a contribution to account for the exact continuum placement in the measurement and the fact that real stars may not be accurately described by a rotational profile ($\pm0.005$). We chose all these error contributions conservatively and add them directly (not in quadrature) because they have a systematic character with exception of the measurement scatter.
We were unable to repeat this procedure for the $R$ band, since we found no combination of orders for which the light ratio corresponds to the $R$ band light ratio.

\subsection{Spectroscopic analysis: $T_{\rm eff}$ and $[\mathrm{Fe/H}]$}
\label{sec:metal}

To determine effective temperatures and metallicities of the binary stars, we first separated their spectra. 
This was done following the description of \cite{Gonzalez06}, which we expanded to handle systems with three 
components (needed for V20).
When separating spectra of binary stars the light contribution from each component to the continuum level cannot be 
determined from the spectra themselves (except when the spectral types are almost identical, where one can follow 
the procedure we used for V18 in the previous section).
We therefore employed our photometric light ratios from Tables \ref{tab:v18_ebop}--\ref{tab:v80_ebop} to correct each component spectrum for the continuum contribution from the other component. Since these light ratios are measured in the Johnson $V$ band, we used Planck functions to calculate the corresponding light ratio in each spectral order from initial photometric temperature estimates. This procedure was iterated with the derived temperatures, with negligible differences. 
We checked that the Planck approximation was adequate, e.g. using MARCS models instead produced identical results.
The binary target V80 was observed with an additional star on the slit on some epochs. After acquiring the spectra, a more careful inspection revealed that this was in fact two very close stars, which we also analyse below. 
For simplicity we shall refer to this composite spectrum as the subgiant (SG) star spectrum.

We made a detailed spectral analysis of four of the separated binary spectra,
the primary and secondary of V18 (abbreviated V18p and V18s), V20p, V80p, and SG. 
We did not analyse V20s and V80s, since the signal--to--noise (S/N) was very low, 
and therefore the placement of the continuum introduced very large systematic errors. 
The continuum was normalized using a synthetic template spectrum with the assumed 
atmospheric parameters of the star. This was done by identifying continuum windows
with the {\tt rainbow} software as described in \cite{Bruntt10b}. 

\begin{figure*}
\centering
\includegraphics[width=7.2cm,angle=90]{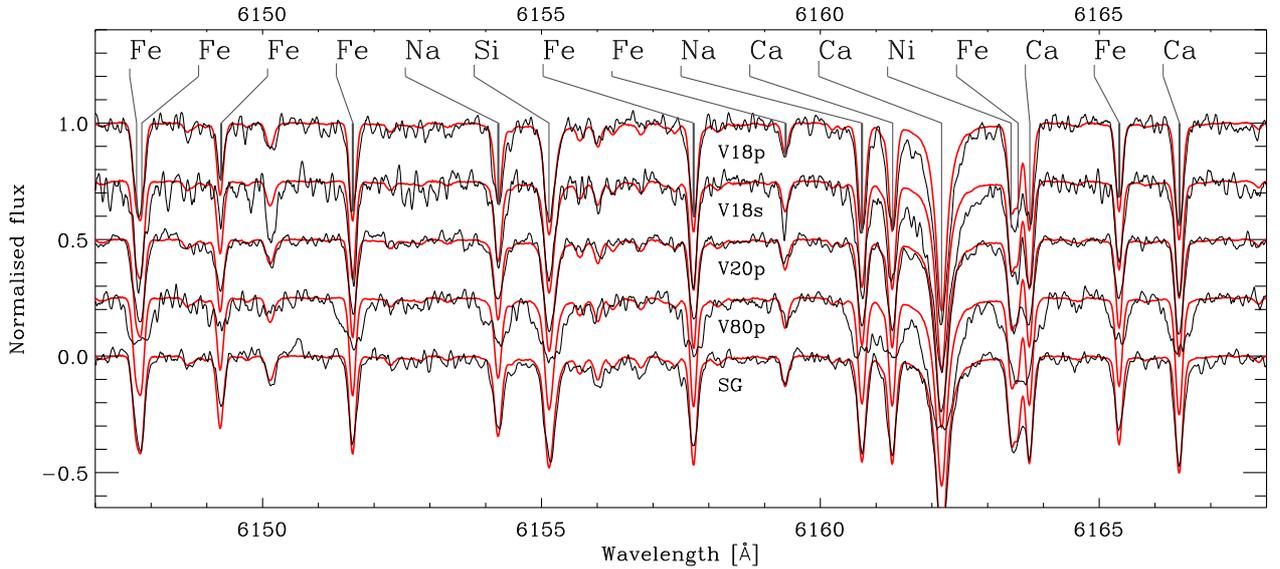}
\caption[]
{
\label{fig:spec}
The five spectra analysed with VWA are compared. The continuum level of the spectra is offset in steps of 0.25 for clarity.
Each spectrum is compared with the HARPS spectrum of $\alpha$~Cen~A (thin red line).
}
\end{figure*}

In Fig.~\ref{fig:spec} we show a small section of the spectra, 
which shows their range of quality. The full-length one-dimensional spectra can be found in Tables 18--22 on CDS. 
The S/N in the continuum ranges from 30 (V18s) to 60 (V20p).
In Fig.~\ref{fig:spec} we also overplot a high resolution (R=110000) high S/N spectrum of $\alpha$~Cen~A\ as observed with HARPS.
This star has parameters very similar to the observed stars in NGC~6791: 
\cite{Bruntt10a} used the HARPS spectrum to determine 
$v \sin i$ $= 1.9\pm0.6$ km/s,
$T_{\rm eff} = 5745\pm80$~K, 
$\log g = 4.31\pm0.08$,
and ${\rm [Fe/H]} = +0.22\pm0.07$.

We used the semi-automatic software package VWA \citep{Bruntt10a, jvc08-vwa} to analyse the spectra.
The program interpolates in a grid of MARCS model atmospheres \citep{gustafsson08}
and uses atomic line data from VALD \citep{kupka99}. The spectral lines are fitted
iteratively by calculating synthetic profiles using SYNTH \citep{valenti96}. Oscillator strengths are corrected using a solar spectrum. The software is described more thoroughly in \cite{Bruntt10a, Bruntt10b}, and we will 
mention here only differences to the standard approach.
Most importantly, we made the analysis by fixing the $\log g$\ values as constrained 
by the accurate measurements of the eclipsing binaries (Table \ref{tab:absdim}). 
This breaks the degeneracy between $\log g$\ and $T_{\rm eff}$, which is usually a problem when analysing spectra at this S/N level, and allows a reliable $T_{\rm eff}$\ determination.

As discussed above, the SG spectrum is a composite of the spectra of two stars with nearly identical
radial velocity. We identified them to be stars 1621 and 1627 in the photometry of SBG03 (finding chart in Fig.~\ref{fig:V80fc}), but their measured magnitudes are wrong due to their small separation. We therefore transformed their magnitudes from our {\it HST} photometry into $B$ and $V$ magnitudes from which we measured their $V$ light ratio and extrapolated along an isochrone to obtain their $\log g$\ values (Table~\ref{tab:ab}). Since they are very close to the binaries in the CMD, this should give $\log g$\ values with errors of less than 0.05 dex. The spectrum of SG was then analysed using the approach for composite spectra described in \cite{jvc08-vwa}. As for the binaries our prior knowledge of $\log g$\ of the two components made it possible to constrain their temperatures.

We adjusted the $v \sin i$\ of the synthetic spectra to provide the best fit to the observations. The values of $v \sin i$\ are listed in Table~\ref{tab:ab}. We used the calibration from \cite{Bruntt10a} to fix the macroturbulence at $2.0$ km/s for all stars. Changing the macroturbulence within reasonable errors does not change the measured equivalent widths, and thus does not affect the results of the spectroscopic analysis. The microturbulence could only be determined with rather large errors. However, the values found are consistent with the calibration for solar-type stars from \cite{Bruntt10a}.

For all the analysed spectra, we adjusted the $T_{\rm eff}$\ of the 
atmosphere model to make the mean abundances of Fe\,{\sc i}\ and Fe\,{\sc ii}\ lines agree. 
We evaluated the internal error on $T_{\rm eff}$\ by changing its value until the
Fe\,{\sc i}\ and Fe\,{\sc ii}\ deviated by one sigma. 

 To confirm the method, \cite{Bruntt10a} compared the $T_{\rm eff}$s from VWA 
 with results for 10 nearby stars where interferometric methods can be used,
 and found very good agreement. They claimed a systematic offset of 
 $T_{\rm eff; VWA} - T_{\rm eff; Interf.} = 40\pm20$~K should 
 be applied to the spectroscopic effective temperatures. 
 However comparing the $(b-y)$ temperature calibration of \cite{casa2010} for 17 stars in common 
 with the spectroscopic analysis of \cite{Bruntt10a}, we find that the situation is more complicated. 
 The derived temperature offset is sensitive to the absolute magnitude of Vega, which is difficult to determine (see \citealt{casa2010} for a discussion). Equally importantly, there are indications that one cannot interpret the $T_{\rm eff}$\ difference between VWA and other methods as a simple offset, since differences may depend on both $T_{\rm eff}$\ and $[\mathrm{Fe/H}]$. Unfortunately, we do not have enough stars measured to reach a firm conclusion. Due to these complications, any attempt to apply an offset to the $T_{\rm eff}$\ derived using VWA may introduce more systematic error. Therefore, we adopt the $T_{\rm eff}$\ values as determined from VWA analysis without any correction and add in quadrature a systematic error of 70 K to the internal VWA errors.

In Table~\ref{tab:ab} we list the adopted $\log g$, the determined $T_{\rm eff}$, $v \sin i$, microturbulent velocity, and $[\mathrm{Fe/H}]$\ for all stars analysed. The errors for $[\mathrm{Fe/H}]$\ are determined by changing the $T_{\rm eff}$\ and microturbulent velocity by one sigma in the analysis. The quoted errors in this table are internal errors.

With the final atmospheric parameters determined, we computed the abundances of all elements with at least four unblended lines available. In Table~\ref{tab:spec} we list the mean abundances and number of lines used for each spectrum. We only used relatively weak
lines with equivalent widths between 10 and 100 m\AA. The quoted errors are the rms of the abundances for each element in a given ionization stage. All elements are consistent within 0.1 dex, except for vanadium, which has a higher mean abundance in all stars.

We tested the effect of the assumed helium content, following the prescription of \cite{stromgren82}. They argue that a
change in the helium content is effectively the same as a change in the pressure, and 
hence $\log g$, in the atmosphere. Using their prescription, we find that even a large change in $Y$ from solar to $0.33$ (high value for NGC~6791) is accommodated by a change in $\log g$\ of only 0.03 which leads to a change in $[\mathrm{Fe/H}]$\ of 0.01 or less. At the same time, helium diffusion would tend to lower the helium content in stellar atmospheres with time. Thus, for the old  stars in NGC~6791 more helium will have left the atmospheres due to diffusion compared to the Sun, and therefore the correction due to a different helium abundance should be even smaller and effectively negligible.
Finally, we repeated the analysis of the stars when changing the light ratio by
two sigma, and found negligible changes to the determined $T_{\rm eff}$\ and metallicity. 

For our final estimate of $[\mathrm{Fe/H}]$\ for NGC~6791 we calculated the weighted mean of the six stars [Fe/H] in Table~\ref{tab:ab}, resulting in ${\rm [Fe/H]_{NGC~6791}} = +0.29 \pm 0.03 \pm 0.07$, where the quoted uncertainties are the weighted  
mean error and the adopted systematic error.

There are several spectroscopic measurements of the metallicity of NGC~6791 in the literature, with results spanning a rather wide range in $[\mathrm{Fe/H}]$\ from $+0.30$ to +$0.47$. Nearly all are all based on giant stars \citep{friel93, peterson98, gratton06, Origlia06, carraro06}. These analyses of giant stars are complicated by the presence of molecules and blending of lines,
especially since they are based on low-to-medium resolution spectra (all have $R<30\,000$). 
In such cases, blending makes the placement of the continuum 
difficult (cf.\ Fig.~2 in \citealt{gratton06} and Fig.~2 in \citealt{carraro06}). 
\cite{Boesgaard09} analysed two sub-giant stars at higher resolution ($R=45\,000$), 
but they adopted a $\log g$\ which is 0.3 dex higher than inferred from our binary measurements. Interestingly, their different $\log g$\ and $T_{\rm eff}$\ parameters conspire to give ${\rm [Fe/H]} = +0.30\pm0.08$ \citep{Boesgaard09}, in agreement with our result, despite their different atmospheric parameters. 

We believe that the metallicity determined from our sample of stars is reliable,
since the parameters of the stars are close to the Sun, and therefore
less prone to systematic effects coming from the adopted model atmospheres. Furthermore, our use of accurate $\log g$\ values from the binary measurements has strongly constrained one free parameter in the spectroscopic analysis. We note also that the $[\mathrm{Fe/H}]$\ measurements of the two single stars in the SG spectrum give results in excellent agreement with the binary stars. This is clear evidence that the spectral separation of the binary stars has not affected the metallicity measurements.   
 
\begin{table*}
 \centering
 \caption{\label{tab:ab}Parameters determined from the spectral analysis ($\log g$ values were fixed). The uncertainties are internal errors.
}
 \setlength{\tabcolsep}{3pt} 
\begin{tabular}{l|llllll}
\hline
\hline\noalign{\smallskip} 
          & V18p             &  V18s            &   V20p           &  V80p             &   SGp           &  SGs           \\ \hline
$T_{\rm eff}$\ [K]& $ 5600 \pm   50$ & $ 5430 \pm 95  $ & $ 5645 \pm  50 $ & $ 5600 \pm 95  $  &  $ 5540\pm60  $ & $ 5570\pm75  $ \\
$\log g$     & $ 4.35         $ & $ 4.43         $ & $ 4.18         $ & $ 4.21         $  &  $ 4.04       $ & $ 4.19       $ \\
$v \sin i$    & $ 4.00 \pm 1.00$ & $ 3.50 \pm 1.00$ & $ 4.50 \pm 1.00$ & $ 14.00 \pm 2.00$  &  $ 4.00\pm1.00$ & $ 4.00\pm1.00$ \\
$v_{\rm micro}$   & $ 1.00 \pm 0.12$ & $ 0.95 \pm 0.10$ & $ 0.90 \pm 0.10$ & $ 1.10 \pm 0.10$  &  $ 1.20\pm0.12$ & $ 1.25\pm0.14$ \\
$[\mathrm{Fe/H}]$      & $+0.31 \pm 0.06$ & $+0.22 \pm 0.10$ & $+0.26 \pm 0.06$ & $+0.34 \pm 0.10$  &  $+0.32\pm0.07$ & $+0.30\pm0.08$ \\ 
\noalign{\smallskip}
\hline

\end{tabular}
\end{table*}

\begin{table*}
 \centering
 \caption{Abundances and number of lines used in the spectral analysis. The uncertainties are the standard deviation of the mean.
 \label{tab:spec}}
 \setlength{\tabcolsep}{3pt} 
\begin{tabular}{l|lr|lr|lr|lr|lr|lr}
\hline
\hline\noalign{\smallskip} 

           & \multicolumn{2}{c|}{V18p} & \multicolumn{2}{c|}{V18s} & \multicolumn{2}{c|}{V20p} & \multicolumn{2}{c|}{V80p} &  \multicolumn{2}{c|}{SGp} & \multicolumn{2}{c}{SGs}  \\
\hline
  {Si \sc   i} &  $ +0.35\pm0.05$  &   9 &  $ +0.46\pm0.06$  &   4 &  $ +0.23\pm0.03$  &  10 &  $ +0.32\pm0.04$  &   6 &                     &     &                   &      \\
  {Ti \sc   i} &  $ +0.26\pm0.04$  &  11 &  $ +0.28\pm0.05$  &   6 &  $ +0.27\pm0.06$  &   7 &                   &     &    $ +0.17\pm0.04$  &   8 &  $ +0.11\pm0.07$  &   8  \\ 
  {V  \sc   i} &  $ +0.48\pm0.07$  &   7 &  $ +0.58\pm0.05$  &   8 &  $ +0.35\pm0.03$  &   8 &  $ +0.57\pm0.07$  &   5 &                     &     &                   &      \\ 
  {Cr \sc   i} &  $ +0.11\pm0.05$  &   4 &                   &     &                   &     &      $-$          &     &                     &     &                   &      \\ 
  {Fe \sc   i} &  $ +0.31\pm0.01$  &  63 &  $ +0.22\pm0.03$  &  50 &  $ +0.26\pm0.01$  &  76 &  $ +0.34\pm0.03$  &  30 &    $ +0.32\pm0.02$  &  56 &  $ +0.30\pm0.02$  &  56  \\ 
  {Fe \sc  ii} &  $ +0.31\pm0.04$  &   5 &  $ +0.21\pm0.04$  &   5 &  $ +0.26\pm0.08$  &   6 &  $ +0.34\pm0.05$  &   5 &    $ +0.31\pm0.02$  &   6 &  $ +0.31\pm0.03$  &   6  \\ 
  {Ni \sc   i} &  $ +0.38\pm0.03$  &  17 &  $ +0.35\pm0.05$  &  18 &  $ +0.31\pm0.02$  &  18 &  $ +0.42\pm0.06$  &  10 &    $ +0.42\pm0.05$  &  10 &  $ +0.34\pm0.04$  &  10  \\ 
\noalign{\smallskip} 
\hline
\end{tabular}
\end{table*}

\section{Absolute dimensions and distance}
\label{sec:absdim}

\subsection{The reddening of NGC\,~6791}
\label{sec:reddening}

The reddening, $E(B-V)$, is needed to obtain the true distance of the cluster, and to constrain isochrones in the CMD. Using V18p, V18s, and V20p, we restricted the acceptable range of $E(B-V)$ by requiring our spectroscopic values of $T_{\rm eff}$\ to agree with the temperatures obtained from a photometric colour-temperature calibration. We used the calibrations from \cite{casa2010} and from MARCS model atmospheres (calculated for use in \citealt{VandenBerg10}). For a range of $E(B-V)$ we found the de-reddened colours of the binary components, and from those the photometric temperatures, assuming $[\mathrm{Fe/H}]$\ = $+0.3$. The two calibrations give very consistent $T_{\rm eff}$s with the largest deviation being 40 K. We set the one-sigma limits on $E(B-V)$ where differences between the spectroscopic and the photometric $T_{\rm eff}$s become greater than 200 K for one of the stars, or greater than 100 K for all three stars. This one sigma range in $E(B-V)$ is 0.135--0.185 with the best match at 0.16. Repeating the procedure for an assumed $[\mathrm{Fe/H}]$\ of $+0.2$ and $+0.4$ showed that the result is only slightly dependent on $[\mathrm{Fe/H}]$\ and thus the acceptable range in $E(B-V)$ is independent of $[\mathrm{Fe/H}]$\ within $\pm0.1$ dex of our $[\mathrm{Fe/H}]$\ measurement. We thus find from this method that $E(B-V) = 0.160\pm0.025$ in excellent agreement with other recent results (VandenBerg et al. 2010; and references therein), and our other measurement, which we now describe.
For an additional measure of the interstellar reddening, we used the calibration of \cite{Munari97} for the relationship between the equivalent width (W) of interstellar NaI D lines and $E(B-V)$. We first combined all our binary spectra at their barycentric velocity in order to average out the spectral signal from the stars. In the combined spectrum the continuum has been suppressed to the left of the interstellar lines because of the NaI D lines from the binary stars, which are at different wavelengths in each spectrum, because the stars orbit each other. We assumed that the continuum suppression above a line across the interstellar NaI D line removes the stellar contribution to the interstellar line depth. Removing first this contribution, we measured the equivalent widths of the interstellar NaI D lines. We repeated the procedure for different selections of spectra, first all binary spectra, then selecting only spectra where deep lines are as far from the interstellar line as possible, and then this same procedure using only spectra from one star.

\begin{table*}   
\begin{center}
\caption[]{\label{tab:reddening}
Equivalent width measurements of the interstellar NaI D lines.
}
\begin{tabular}{lcccc} \hline    
\hline
\noalign{\smallskip}    
Spectra	used		&	W(D1)	&	W(D2)	& W(D1) min. from W(D2)	&	W(D1) from W(D1)+W(D2)\\
\hline    
\noalign{\smallskip}    
V18+V80+V20		& 0.3406	& 0.3200	&	0.3520	&	0.346\\
(V18+V80+V20)selected	& 0.3544	& 0.3222	&	0.3544	&	0.354\\
V80selected		& 0.3309	& 0.3558	&	0.3913	&	0.361\\
\noalign{\smallskip}    
\hline    
\end{tabular}            
\end{center}
\tablefoot{
W(D1) and W(D2) are direct measurements of the NaI lines. W(D1) min. from W(D2) is a minimum value of W(D1) calculated from W(D1)/W(D2)\,$\geq1.1$. W(D1) from W(D1)+W(D2) is the mean value of the two estimates of W(D1).}
\end{table*}
                                  
Munari \& Zwitter (1997) find that W(D1)/W(D2)$\geq$ 1.1, which enables a minimum estimate of W(D1) from W(D2).
Since in our case this minimum estimate is slightly higher than the direct measurement, we use the mean of the two measurements as our final measurement. Results are shown in Table~\ref{tab:reddening} from which we adopt $0.354\pm0.010$ as the measured W of the NaI D1 line giving most weight to the measurement using many binary spectra while avoiding strong stellar contamination. Using this value with Table 2 of Munari \& Zwitter (1997) we find $E(B-V)=0.153\pm0.010$. Munari \& Zwitter state that for $E(B-V)$ below 0.5 the reddening uncertainty is $\sim$ 0.05, but from their Fig. 5 it is clear that for $E(B-V)$ below 0.2, the uncertainty is even smaller. Despite this we conservatively adopt $E(B-V)=0.153\pm0.050$ as our measurement from this method, which is admittedly not very precise but in excellent agreement with the above result.

We adopt the most accurate of our two measurements to be used in the following:
$E(B-V) = 0.160\pm0.025$.

\subsection{Absolute dimensions and distance of V18, V20 and V80}
\label{sec:absdim}

Absolute dimensions for V18, V20, and V80 are calculated
from the elements given in Tables~\ref{tab:v18_ebop}--\ref{tab:orbit}, and listed in Table~\ref{tab:absdim}.
As seen, masses and radii have been
obtained to an accuracy well below 1\% for both V18 and V20. Errors on the masses are in the range 0.27--0.36\%, while errors on the radii are in the range 0.61--0.92\%. These measurements are therefore of an accuracy comparable to, or better than, that of field star eclipsing binaries \citep{Torres10}.
The masses of V80 are also measured with errors below 1\%, but as expected the radii are not very precise. However, the error in $\log g$\ for the primary component is still only $\pm 0.05$, which was useful for the $[\mathrm{Fe/H}]$\ measurement.  
Our measurements of the radius of the V20 secondary and the mass of the V20 primary are outside the $1\,\sigma$ error estimates of GCH08. These differences are due to our improved measurements of the third light and radial velocity zero-points.

As also seen in Table~\ref{tab:absdim}, our measurements of the rotational velocities are in agreement with the theoretical synchronous rotational velocities. V18 has an eccentric orbit, but the difference between theoretical synchronous and pseudo-synchronous velocities is only 0.1 km/s and has been included in the theoretical values by increasing the error of the theoretical prediction. 

From our measured radii and spectroscopic $T_{\rm eff}$\ values, we calculated the luminosities of V18p, V18s and V20p, which together with a bolometric correction (B.C.) gave $M_V$ of the stars. From these, and our measured $V$ and $E(B-V)$, we calculated the distance moduli and true distance, which we also tabulate in Table~\ref{tab:absdim}. As seen the distance moduli of the three components agree very well, with the error in the $T_{\rm eff}$ values being the dominant error contribution. We used B.C. from MARCS stellar atmospheres, which are in agreement with empirical B.C.s from Flower (1996) within 0.02. The true distance moduli were found using our measurement of $E(B-V) = 0.160\pm0.025$ and $A_V=3.09 \times E(B-V)$ \citep{McCall04}. From the individual distance moduli of V18p, V18s, and V20p, we calculated the apparent and true distance moduli of NGC~6791 to be $(m-M)_V = 13.51 \pm0.06$ and $(V_0-M_V) = 13.01 \pm0.08$. The distance is then d(NGC~6791) = $4.01\pm0.14$ kpc.

\begin{table*}   
\begin{center}
\caption[]{\label{tab:absdim}
Astrophysical data for V18, V20 and V80.  
}
\begin{tabular}{lrrrrrr} \hline    
\hline    
\noalign{\smallskip}    
                     &    \multicolumn{2}{c}{V18}       &    \multicolumn{2}{c}{V20}  & \multicolumn{2}{c}{V80}\\ 
\noalign{\smallskip}    
                     &  \multicolumn{1}{c}{Primary}   & \multicolumn{1}{c}{Secondary} & \multicolumn{1}{c}{Primary} &    \multicolumn{1}{c}{Secondary} & \multicolumn{1}{c}{Primary} &    \multicolumn{1}{c}{Secondary}     \\ 
\noalign{\smallskip}    
\hline    
\noalign{\smallskip}    
Absolute dimensions: &                  & &&                 \\ 
$M/M_{\sun}$     &$0.9955 \pm 0.0033$ &$0.9293 \pm 0.0032$ &$1.0868 \pm 0.0039$ &$0.8276 \pm 0.0022$ &$1.0588 \pm 0.0091$ &$0.8003 \pm 0.0062$
\\ 
$R/R_{\sun}$     &$1.1011 \pm 0.0068$ &$0.9708 \pm 0.0089$ &$1.397 \pm 0.013$  &$0.7813 \pm 0.0053$ &$1.341 \pm 0.081$ &$0.90 \pm 0.18$
\\ 
$\log g$ (cgs)   &$4.3524 \pm 0.0053$ &$4.4319 \pm 0.0080$ &$4.1840 \pm 0.0078$ & $4.5698 \pm 0.0059$ &$4.208 \pm 0.052$ & $4.43 \pm 0.18$
\\ 
\\
$T_{\mbox{\scriptsize eff}}\,$ (K)\tablefootmark{a}  &  $5600 \pm 95$ &   $5430\pm 125$ &   $5645 \pm 95$ & $4824 \pm 140$    &   $5600 \pm 95$ & $(...)$   \\
\\
$v_{rot}$ (km\thinspace s$^{-1}$)\tablefootmark{b}& $ 4.0 \pm 1.0$    & $ 3.5 \pm 1.0$ & $ 4.50 \pm 1.00$    &  $(...)$  & $ 14.1 \pm 2.0$    & $(...)$ \\
$v_{sync}$ (km\thinspace s$^{-1}$)\tablefootmark{b}& $ 2.96 \pm 0.14$    & $ 2.61 \pm 0.13$ & $ 4.88 \pm 0.04$    & $ 2.73 \pm 0.02$ & $ 13.88 \pm 0.83$    & $ 9.33 \pm 1.9$\\
\\            
$V$				    & $18.247\pm0.020$	& $18.740\pm0.020$	& $17.695\pm0.020$ & $19.874 \pm 0.020$ & $(...)$\tablefootmark{c} & $(...)$\tablefootmark{c}\\
$M_V$				    & $4.76\pm0.10$     & $5.22\pm0.10$         & $4.19\pm0.10$ & $6.44\pm0.18$ &$(...)$\tablefootmark{c}& $(...)$\tablefootmark{c}\\
$(V-M_V)$			    & $13.49\pm0.10$	& $13.52\pm0.10$	& $13.50\pm0.10$ & $13.44\pm0.18$& $(...)$\tablefootmark{c}&$(...)$\tablefootmark{c}\\
$(V_0-M_V)$			    & $13.00\pm0.13$	& $13.03\pm0.13$	& $13.01\pm0.13$ & $12.95\pm0.20$& $(...)$\tablefootmark{c}&$(...)$\tablefootmark{c}\\
distance (kpc)			    & $3.98\pm0.25$	& $4.04\pm0.25$	& $4.00\pm0.25$ & $3.89\pm0.34$ & $(...)$\tablefootmark{c}&$(...)$\tablefootmark{c}\\
\noalign{\smallskip}            
\hline
\end{tabular}
\end{center}            
\tablefoot{}
\tablefoottext{a}{$T_{\rm eff}$\ values are our spectroscopic measurements, except for $T_{\rm eff}$\ of V20s, which is calculated as $T_{\rm eff}$\ of V20p minus the difference in photometrically predicted temperatures between V20p and V20s assuming $E(B-V)$=0.16.}
\tablefoottext{b}{
$v_{rot}$ is the observed equatorial rotational velocities calculated by combining $v \sin i$\ and $\sin i$\ measurements. $v_{sync}$ is the theoretical equatorial velocity for synchronous rotation.
}
\tablefoottext{c}{
Absolute magnitudes and distance moduli have not been calculated for V80 because these are very uncertain due to uncertain $V$ magnitudes and radii (and $T_{\rm eff}$\ for V80s) caused by the magnetic activity.
}
\end{table*}                   
\begin{figure}
\epsfxsize=80mm
\epsfbox{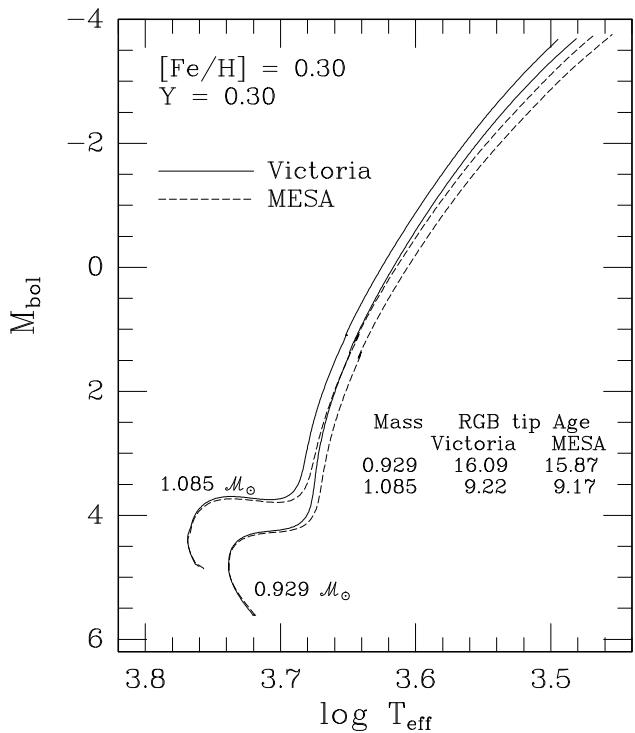}
\caption[]{\label{fig:n6791mesauvic}
Comparison between Victoria and MESA stellar evolutionary tracks for masses
and chemical abundances (as indicated) that are relevant to an investigation
of NGC~6791 and the binary stars V18 and V20.
}
\end{figure}

\section{A first comparison with theoretical models}
\label{sec:modelc}

GCH08 used their measurements of V20 to show that the difference
in its predicted age due to the specific stellar models that were employed was
about four times greater than their measurement precision. However, they were unable to determine which of the models (if any) were the most
trustworthy ones. The present study has significantly improved the measured
properties of the binaries --- including [Fe/H], $T_{\rm eff}$, and $E(B-V)$;
consequently, even tighter constraints may be placed on the models. Of
particular importance is the fact that, given our accurate mass and radius
measurements for both V18 and V20, we can define four points (and their
uncertainties) on the mass-radius diagram and therefore constrain the curvature
of the isochrone that best reproduces the observations.

Although we could repeat the analysis carried out by GCH08, an 
examination of the models that they used revealed that the
different assumed helium mass-fraction, $Y$, of NGC~6791, is
the primary reason for the differences in the ages that they derived
\citep{Brogaard10}.  Since most of the currently available isochrones do not
contain variations in $Y$ at fixed $[\mathrm{Fe/H}]$\ values, a meaningful comparison of the
models produced by different workers cannot be made.  Moreover, there have been
a number of recent revisions to nuclear reaction rates and other physics
ingredients (see the next section), which further limits the usefulness of
comparisons with most published models because they are not based on the most
up--to--date stellar physics. In this paper, we have therefore chosen to make
some initial comparisons with isochrones for a range in $[\mathrm{Fe/H}]$\ and $Y$, based on
the tracks computed using just one stellar evolutionary code that has
incorporated recent advances in nuclear reactions, opacity, etc.

\subsection{Stellar models}
\label{sec:models}

The stellar models that are used in the present and forthcoming analysis of the
binary stars V18 and V20 in NGC$\,$6791, together with the cluster CMD, were
generated using a significantly updated version of the University of Victoria
evolutionary code (last described by \citealt*{VandenBerg06}; and references
provided therein).  The most important improvement that has since been made to
it is the incorporation of the treatment of the gravitational settling of helium
presented by \cite*{Proffitt91}.  A simple prescription for turbulent
mixing has also been implemented, given that additional mixing below envelope
convection zones must be assumed in order for theoretical models to reproduce
the observed variations of the surface metal abundances of stars between the
main sequence and the giant branch in old stellar systems (e.g., see the study
of NGC$\,$6397 by \citealt{Korn07}). The free parameter in the adopted
formulation of turbulence is set by monitoring the settling and nucleosynthesis
of lithium and requiring that a Standard Solar Model yield the observed Li
abundance of the Sun.  In fact, it turns out that, with such a normalization,
models for very metal-poor stars predict variations of the surface helium
abundance with evolutionary state that are quite similar to those reported by
Korn et al. For a much more detailed description of these results and the
adopted treatment of diffusion and turbulence, we refer to a
forthcoming study which examines the impact of varying the abundances of
individual heavy elements (D.~VandenBerg et al., in preparation). The settling of heavy elements is not considered.  However,
as discussed below, this omission is expected to have only minor
consequences for the derived age of NGC$\,$6791.

\begin{figure}
\epsfxsize=80mm
\epsfbox{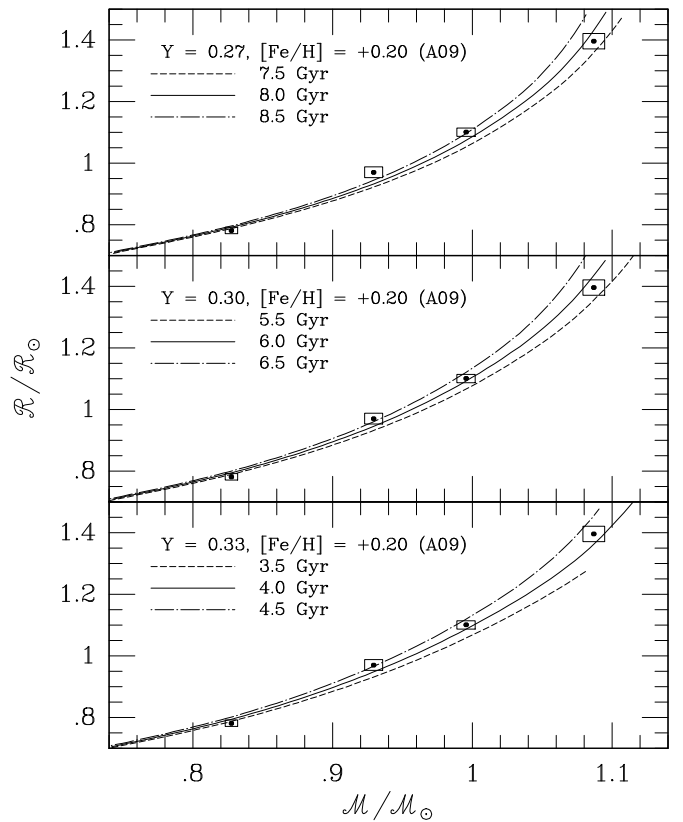}
\caption[]{\label{fig:mrp020}
The measured masses and radii of the components of V18 and V20 compared to best matching isochrones for $[\mathrm{Fe/H}]$\ = $+0.20$ and helium contents of Y = 0.27, 0.30, and 0.33. From left to right the stars are V20s, V18s, V18p, and V20p. Error boxes are two-sigma.
}
\end{figure}

\begin{figure}
\epsfxsize=80mm
\epsfbox{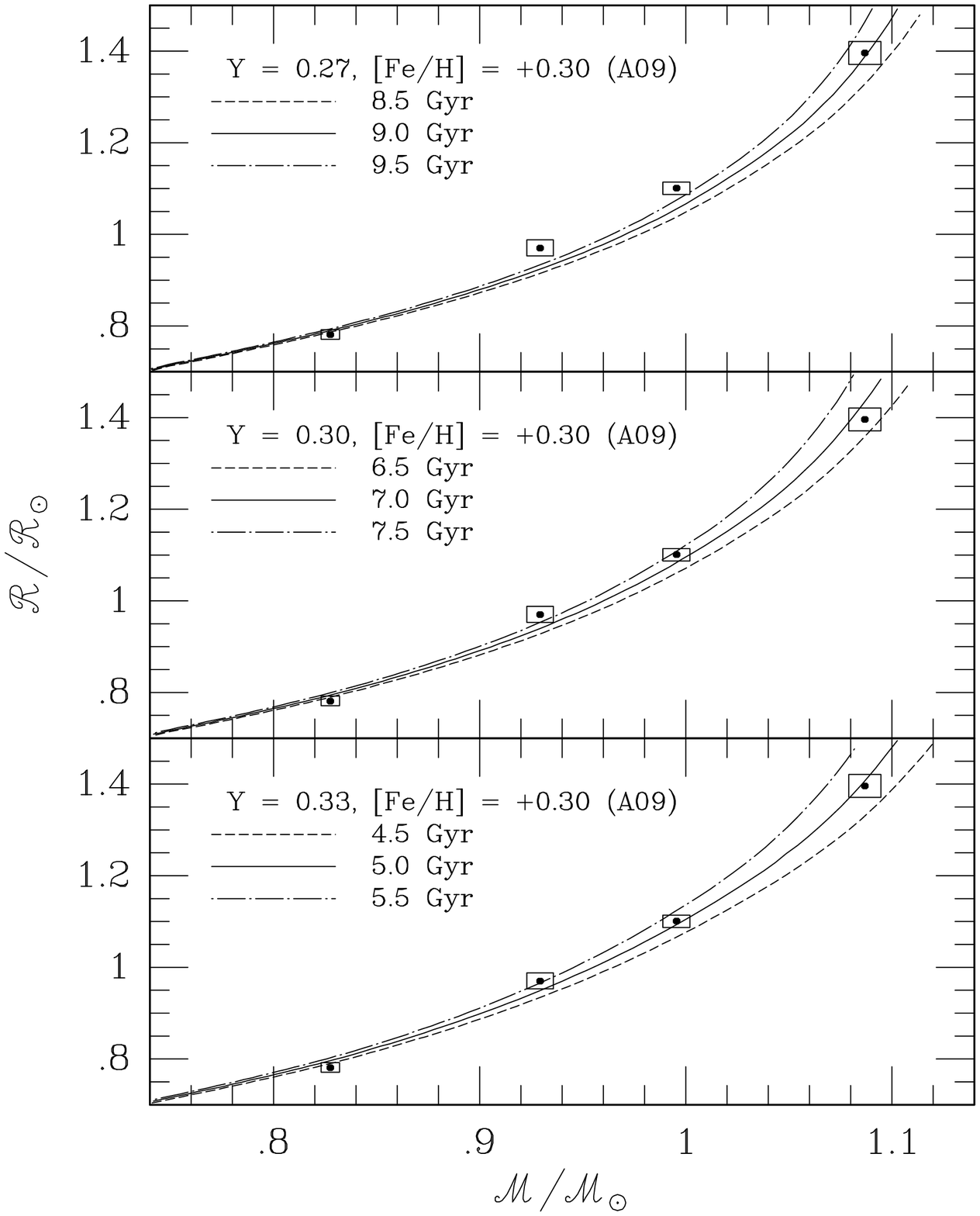}
\caption[]{\label{fig:mrp030}
As Fig.~\ref{fig:mrp020} but for isochrones with $[\mathrm{Fe/H}]$ = $+0.30$
}
\end{figure}

\begin{figure}
\epsfxsize=80mm
\epsfbox{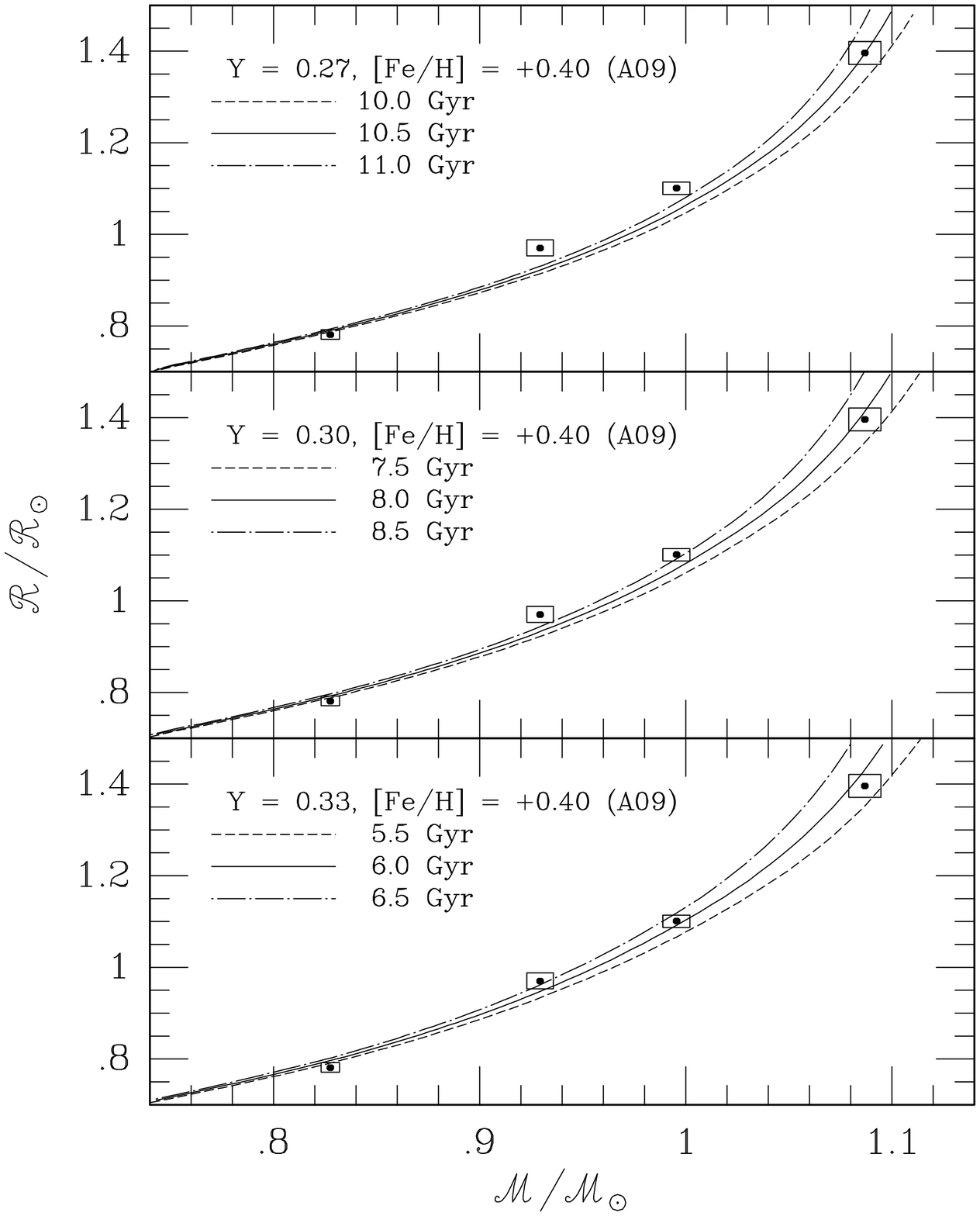}
\caption[]{\label{fig:mrp040}
As Fig.~\ref{fig:mrp020} but for isochrones with $[\mathrm{Fe/H}]$ = $+0.40$
}
\end{figure}

Since 2006, there have been a number of improvements to the rates of several
reactions in the $pp$-chain and the CNO-cycle (for a convenient summary, see
\citealt*{Weiss08}) --- including, in particular, the rate of the ``bottle-neck"
$^{14}$N$(p,\,\gamma)^{15}$O reaction \citep{Marta08}. The current version
of the Victoria code has been revised to take these advances into account, as
well as the improved conductive opacities published by \cite{Cassisi07}.
In addition, particular care has been taken concerning the surface boundary
conditions that are needed to solve the stellar structure equations (see, e.g.,
\citealt{Vandenberg92}), because the predicted $T_{\rm eff}$ scale depends so
sensitively on them.  As demonstrated by \cite{Vandenberg08}, simple
model atmospheres that are obtained by integrating the hydrostatic equation in
conjunction with the scaled empirical \cite*{Holweger74} $T$--$\tau$
relation yield pressures at $T = T_{\rm eff}$ that are in very good agreement
with the values obtained from scaled, differentially corrected MARCS model
atmospheres over wide ranges in $T_{\rm eff}$, $\log\,g$, and metallicity.
That is, the former approach provides a good approximation to the use of modern,
blanketed model atmospheres as boundary conditions; hence, we have opted to
determine the boundary pressures in this way, using the fit to the
Holweger-M\"uller $T$--$\tau$ data given by \cite{VandenBerg89}.  In
support of this choice, \cite*{VandenBerg10} have shown that the resultant
temperatures predicted by stellar models for field subdwarfs are in excellent
agreement with those determined for them using the infra-red flux method.

Fig.~\ref{fig:n6791mesauvic} shows that the Victoria code produces evolutionary
tracks for super-metal-rich (SMR) stars that agree quite well with those
computed for the same masses and initial helium content using the completely
independent MESA code (\citealt{Paxton10}; see also
http://mesa.sourceforge.net). Except for the fact that the MESA code takes
the gravitational settling of the metals into account (i.e., as well as helium)
using the \cite*{Thoul94} formalism, the physics incorporated in the two codes
is quite similar. The MESA computations assume an initial mass-fraction
abundance of the metals, $Z$, that is about 20\% higher than the predicted
surface metallicity at an age of $\sim 7$ Gyr, when the corresponding [Fe/H]
value is $\approx +0.30$.  The surface value of $Z$ in the Victoria
models does not vary with time. Encouragingly, the tracks superimpose one another
nearly exactly between the zero-age main sequence and the lower red-giant
branch (RGB), and they predict the same ages at the tip of the RGB to within
1.4\%, despite the noted differences in the treatment of diffusion. Thus,
models computed using either code can be expected to predict very similar
temperatures and ages for the NGC$\,$6791 binaries considered in this study,
which are all in their core H-burning phases.  Although the settling of the
metals reduces the minimum mass that retains a convective core throughout its
main-sequence phase (see \citealt{Michaud04}), NGC$\,$6791 appears to be old
enough (see below) that this is unlikely to be a concern for the interpretation
of the observations. Indeed, as pointed out by \cite{Christensen-Dalsgaard09}, the $\sim 50$\%
   reduction in the rate of the $^{14}$N$(p,\,\gamma)^{15}$O reaction
   (Marta et al. 2008) has the important consequence of increasing the
   mass of the lowest mass star that retains a convective core throughout
   its main-sequence phase, thereby decreasing the age of the oldest
   isochrone that predicts a gap near the turnoff.
The main reason for the difference in the predicted
location of the RGB is that the MESA code requires a lower value of the
mixing-length parameter ($\alpha_{\rm MLT} \approx 1.8$) than the Victoria code
($\approx 2.0$) to satisfy the solar constraint.

For this initial analysis of NGC$\,$6791 and its binaries, we have adopted the
relative abundances of the heavy elements recently derived for the Sun by
\cite{Asplund09}.

\begin{figure*}
\centering
\includegraphics[width=13cm,angle=270]{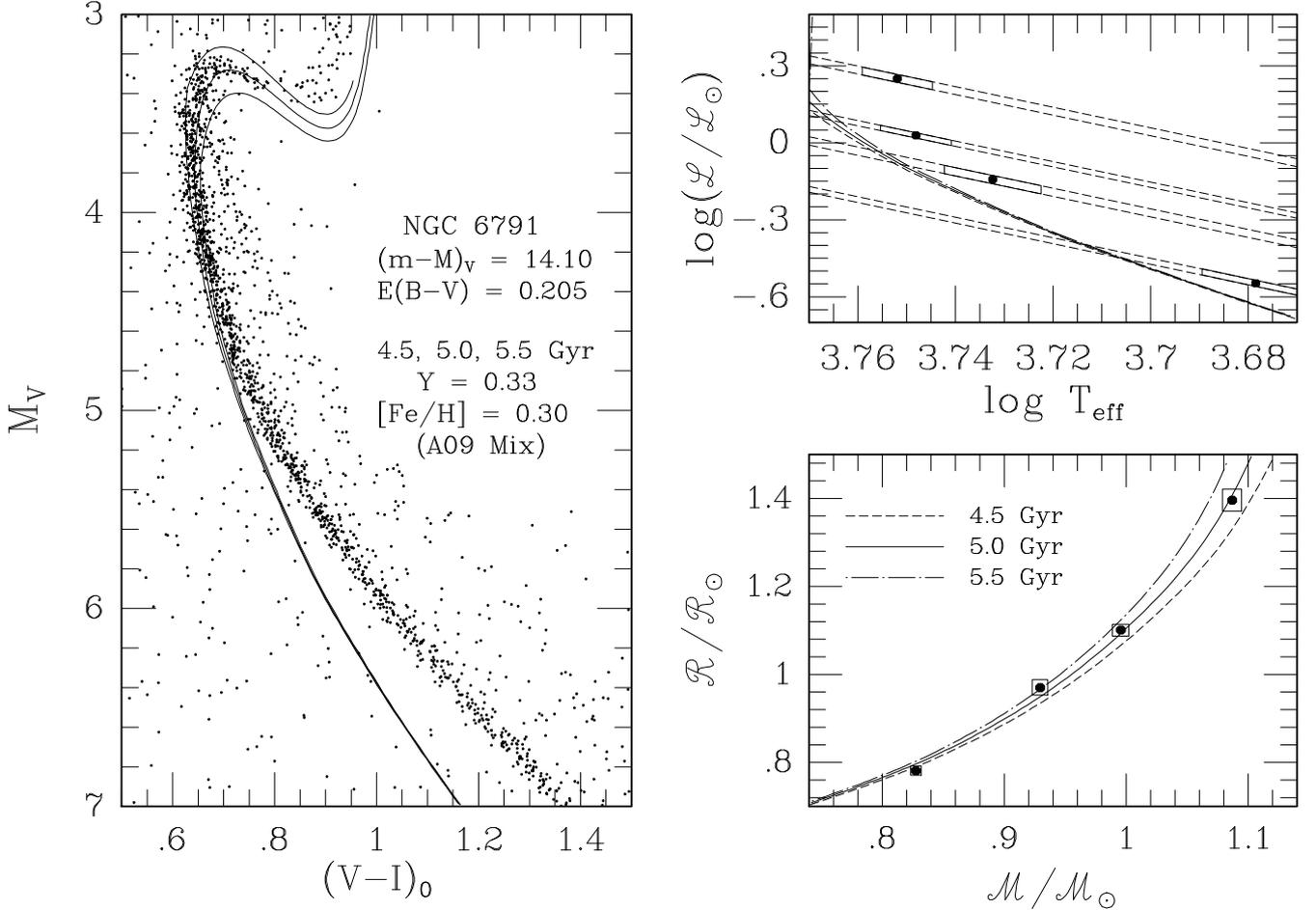}
\caption[]{\label{fig:cmdp030y33}
Measurements of the detached eclipsing binaries V18 and V20 and the cluster $(V-I)_0,V$ CMD compared to isochrones with $[\mathrm{Fe/H}]$ = $+0.30$ and Y = 0.33. The observations in the CMD are from a 2010 re-reduction of the SBG03 data set by P.~Stetson (see \citealt{VandenBerg10}. Only the stars with the highest precision $VI$ photometry have been plotted.
}
\end{figure*}

\begin{figure*}
\centering
\includegraphics[width=13cm,angle=270]{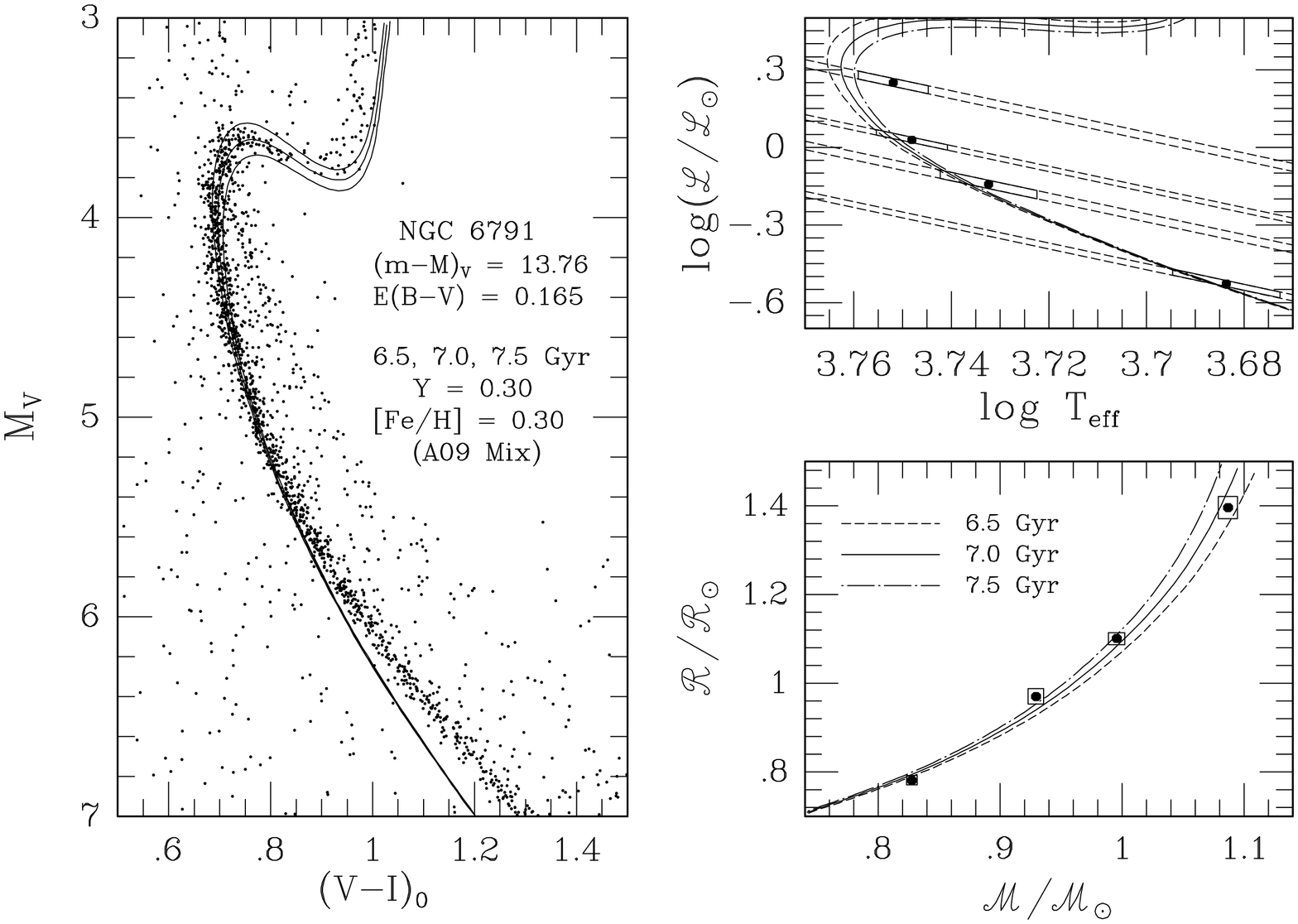}
\caption[]{\label{fig:cmdp030y30}
As Fig.~\ref{fig:cmdp030y33} but for isochrones with $[\mathrm{Fe/H}]$ = $+0.30$ and Y = 0.30
}
\end{figure*}

\subsection{Mass-Radius diagrams}
\label{sec:mr}

The best starting point for an analysis of binary stars is arguably the
mass-radius (MR) diagram because it provides the most direct comparison
between theory and observations that can be made, being independent of
uncertainties in distance, reddening, and colour--temperature transformations.
Since the measured $[\mathrm{Fe/H}]$\ value is very close to $+0.3$ (with a systematic error
that is difficult to determine), we will compare our observations with models
that have $[\mathrm{Fe/H}]$\ $= +0.2$, $+0.3$, and $+0.4$, with $Y = 0.27$, 0.30, and 0.33 at
each $[\mathrm{Fe/H}]$.  In Figs.~\ref{fig:mrp020}--\ref{fig:mrp040} the measured masses and
radii of the components of V18 and V20 are compared with isochrones for these
chemical abundances (as noted) and for the indicated ages, which span the range
needed to match the most massive, and thus most highly evolved, star in the
sample (which is the primary of V20).  It is worth reiterating the fact that
having four measurements in the MR diagram allows us to constrain the shape of
the isochrone on this plane.

The error boxes correspond to two sigma uncertainties.  We have chosen to do
this because, with an increasing number of measurements in the mass-radius
diagram, the likelihood of fitting all of the observations to within one sigma
decreases.  On the other hand, simple statistics cannot be used to predict the
probability of fitting all of the observations to within $1\,\sigma$ (or
$2\,\sigma$) given that errors in the masses and radii can compensate for
one another to a extent that depends on the slope of the isochrone in the
vicinity of each of the observed points.  For more rigorous results, we intend
to use Monte Carlo simulations to quantify how well we should expect to fit
our observations (Brogaard et al. 2010, in prep.; hereafter Paper II).

Figs.~\ref{fig:mrp020}--\ref{fig:mrp040} show quite clearly that the morphology
of an isochrone over the range in mass encompassed by the binaries depends
quite sensitively on the assumed helium abundance (and only weakly on [Fe/H]).
Indeed, it is only because we have measurements of multiple eclipsing binaries
that we are able to use the shape of an isochrone on the MR plane to constrain
the value of $Y$.  This would obviously not have been possible if only V18
(the two middle measurements) or only V20 (the highest and lowest points) were
used in the comparisons with theory --- as a change in $Y$ would then be
indistinguishable from a change in age.  With four stars, this degeneracy is
now broken.  As seen from the three figures, isochrones for $Y = 0.27$ are
unacceptable for all choices of $[\mathrm{Fe/H}]$, because the isochrones miss one star by
more than two sigma, and only touches the others at or near the corners of their
$2\,\sigma$ error boxes.  Isochrones for $Y = 0.30$ match the observations 
somewhat better, though they also tend to pass through the corners of the
$2\,\sigma$ error boxes: the least acceptable of the $Y = 0.30$ isochrones is
that for $[\mathrm{Fe/H}]$ $= +0.40$.  In fact, the shallow curvature suggested by the
measured masses and radii indicates a preference for the highest helium
abundance that we have considered ($Y$ = 0.33), independently of the assumed
$[\mathrm{Fe/H}]$\ value.  However, such a high value of $Y$ requires rather young ages,
which turn out to be problematic, as we will now demonstrate.

\subsection{$T_{\rm eff}$--luminosity and colour-magnitude diagrams}
\label{sec:cmds_6791}

The isochrones plotted in the various MR diagrams that we have considered may,
of course, also be compared with observations of the components of V18 and V20
on the log($T_{\rm eff}$ )--log(L/$\mathrm{L_{\sun}}$)\ (TL) diagram, using luminosities
for the latter which can be readily calculated from their measured radii and the
values of $T_{\rm eff}$\ that were derived in our spectroscopic analysis.  In addition,
the same isochrones may be fitted to the CMD of NGC$\,$6791 to determine the
apparent distance modulus that corresponds to a given age. The value of
$(m-M)_V$ derived in this way should agree with the modulus found from the
binary stars themselves ($13.51 \pm 0.06$, see Table~\ref{tab:absdim}).
An acceptable model must therefore be able to fit the MR and TL diagrams, as
well as the observed CMD, simultaneously.

Fig.~\ref{fig:cmdp030y33} provides an example, for the case of $[\mathrm{Fe/H}]$\ $= +0.30$
and $Y = 0.33$, that can be ruled out because it is not possible for the models
to match the observations on these three planes in a fully consistent way.
Clearly both the predicted effective temperatures and the apparent distance
modulus are way too high.  For the same reasons, the other $Y = 0.33$ cases (as
well as the possibility of even higher helium contents) can be excluded.
Because our examination of the MR diagram has already precluded $Y \le 0.27$ as
viable possibilities, we are then left to consider the models having $Y = 0.30$.
If $[\mathrm{Fe/H}]$\ $= +0.20$, an isochrone for the age that does the best job of
reproducing the MR diagram will fit the turnoff of the cluster CMD only if the
apparent distance modulus is $\sim 13.95$ (not shown), which is unacceptably
high.  As already noted, models for $[\mathrm{Fe/H}]$\ $= +0.40$ provide an unsatisfactory
fit to the mass-radius data, leaving us with only the $Y= 0.30$, $[\mathrm{Fe/H}]$\ $=
+0.30$ case.   As illustrated in Fig.~\ref{fig:cmdp030y30}, these parameter
choices result in a level of agreement between theory and observations that
can be considered satisfactory. The failure of the isochrones to reproduce
the lower main-sequence of NGC$\,$6791 on the CMD could well be due mostly
to problems with the adopted colour--$T_{\rm eff}$\ relations.  The discrepancies would
be significantly smaller had we used the empirical colour transformations given
by Casagrande et al.~(2010); see VandenBerg et al.~(2010). For stars brighter than $M_V \sim 6$ the MARCS and the Casagrande et al. transformations are in good agreement.

While we could explore the impact of small changes in $Y$ and [Fe/H] close to
this solution, we should first expand our comparisons with
theoretical models to those which assume different chemical abundance patterns.
For instance, we have noticed that models which assume the older \cite{Grevesse98}
 metals mixture, which are also favoured from helioseismological
studies (e.g., \citealt{Bahcall05}), are able to reproduce the properties of
the secondary of V20 much better than those that assume the Asplund et al. (2009)
mix of heavy elements.  (Whether better agreement can also be found for the
other binary components remains to be seen.)  In addition, it is worthwhile
to examine the effects of varying the CNO abundances, given that they have
been found to be under-abundant in NGC$\,$6791 by 0.2--0.3 dex relative to scaled
solar \citep{Carretta07, Origlia06} while \cite{attm07} also find indications of C being under-abundant. It is well known that the CNO elements
have important consequences for stars in their core H-burning phases through
both nucleosynthesis and opacity effects. In Paper II we
will present a thorough investigation of these and other issues, in the hope
that we will obtain much more satisfactory fits of stellar models to both the
CMD of NGC$\,$6791 and the cluster binaries than those reported here. Thus,
our best estimates of the age and helium abundance of NGC$\,$6791 will be
reported in Paper II.

\section{Summary and conclusions}
\label{sec:concl}

In this paper we have presented extensive photometric and spectroscopic 
observations of the eclipsing binaries V18, V20 and V80, which are members 
of the old open cluster NGC\,6791, and determined accurate masses and
radii for the components of two of these systems. Additionally we exploited
the eclipsing binaries for reliable measurements of spectroscopic
$T_{\rm eff}$s, and the metallicity and reddening of NGC~6791. 

By performing a combined stellar model comparison with {\it multiple}
eclipsing binaries in MR and TL diagrams and the cluster CMD we showed
that we can constrain stellar models better than ever before. This
allowed us to constrain the helium content, and thereby obtain a more
precise (and hopefully accurate) age of the cluster, although in a
model-dependent way. It turned out that in order for models to match
both the MR and TL diagram and the CMD, only an age close to 7 Gyr is
allowed by the observations, even though models are selected with a range
of $\pm0.1$ dex in $[\mathrm{Fe/H}]$\ and a large range in Y. Our best current
estimate of the cluster age and helium content is $7.0$ Gyr and Y=0.30.
This solution indicates that the helium enrichment law is
$\Delta Y/\Delta Z \sim\, $2, assuming that such a relation exist and
applies to NGC~6791. However, these conclusions may depend on the
details of the stellar models and the detailed abundance pattern of
NGC~6791, which will be investigated in the forthcoming paper II.

We would like to point out that NGC~6791 contains a large number of additional detached
eclipsing binaries \citep{Rucinski96, deMarchi07, Mochejska05}. Accurate
measurements of these, which we have shown to be possible, would allow
even tighter model constraints in the MR diagram by extending the mass range
over which the observed mass--radius relation must be reproduced. This
has the potential to strengthen tests of other physical aspects of
stellar models. This is so because incorrect model physics may be
concealed in the models due to the flexibility offered by the
uncertain helium content, but only up to a certain level. Furthermore,
NGC~6791 is in the Field-Of-View of the NASA {\it Kepler} mission \citep{Borucki10}, which will not
only allow many more detached eclipsing binaries to be found,
and their periods determined, but also complementary model
constraints from asteroseismology of the giant stars in the cluster. 

Multiple detached eclipsing binaries have also been confirmed in a
number of other old open clusters \citep{Brogaard10,Talamantes10}.
Extending this kind of analysis to these clusters will allow more
reliable cluster ages to be determined and more aspects of stellar
models to be tested, when stellar models have to reproduce observations
in clusters with a wide range in age and metallicity. By extending studies
also to the globular clusters, detached eclipsing binaries could
ultimately provide the strongest constraints on their ages and helium
contents.

\begin{acknowledgements}
We thank the staff at the Nordic Optical Telescope and at ESO for
allowing very flexible scheduling of observations, which resulted in optimal
data and a faster completion of this project than would otherwise have
been possible.  We thank J. Southworth for access to JKTEBOP and
P. Stetson for sharing his excellent photometry software with us and for permission to use his latest photometry of NGC6791.  We are also
grateful to Aaron Dotter for computing the MESA evolutionary tracks in
Fig.~\ref{fig:n6791mesauvic}.
The project "Stars: Central engines of the evolution of the Universe",
carried out at University of Aarhus and Copenhagen University, is supported
by the Danish National Science Research Council.
FG acknowledges financial support from the Carlsberg Foundation, the
Danish AsteroSeismology Centre at the University of Aarhus, and 
Instrumentcenter for Dansk Astronomi. 
The following internet-based resources were used in research for
this paper: the NASA Astrophysics Data System; the SIMBAD database
and the ViziR service operated by CDS, Strasbourg, France; the
ar$\chi$iv scientific paper preprint service operated by Cornell University.
\end{acknowledgements}

\bibliographystyle{aa}
\bibliography{15503} 



\appendix
\section{Finding charts}
\label{sec:concl}

\begin{figure*}
\centering
\epsfxsize=80mm
\epsfbox{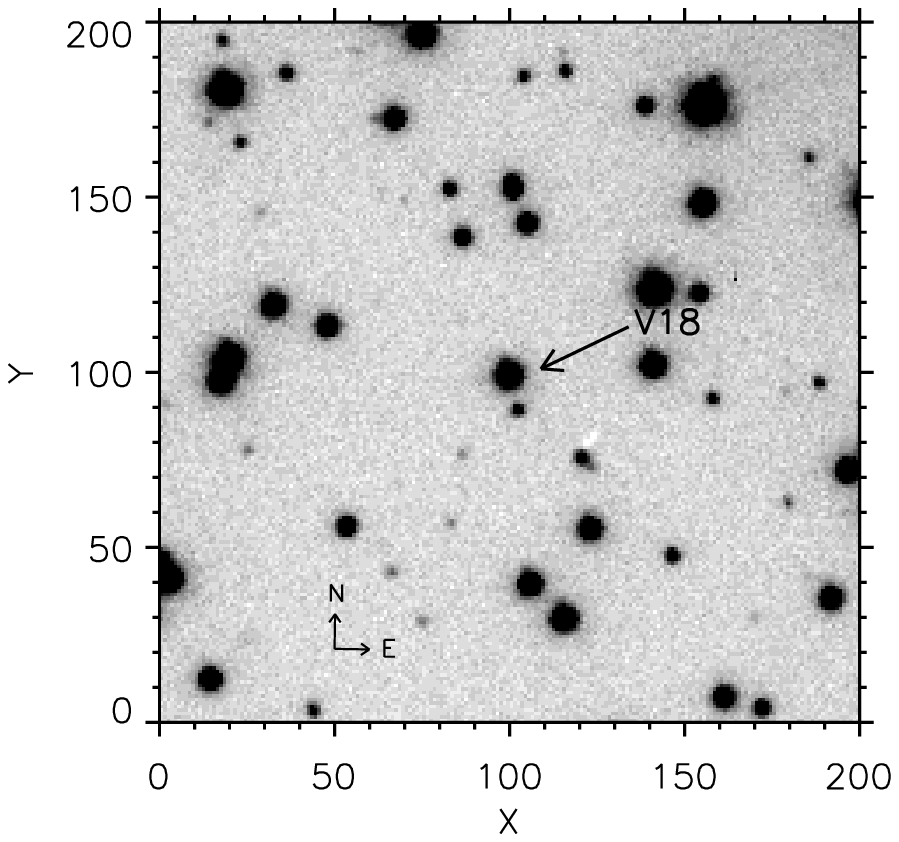}
\caption[]{\label{fig:V18fc}
Finding chart for V18. Plate scale is 0.19\arcsec per pixel. Image from ALFOSC at the Nordic Optical Telescope.}
\end{figure*}

\begin{figure*}
\centering
\epsfxsize=80mm
\epsfbox{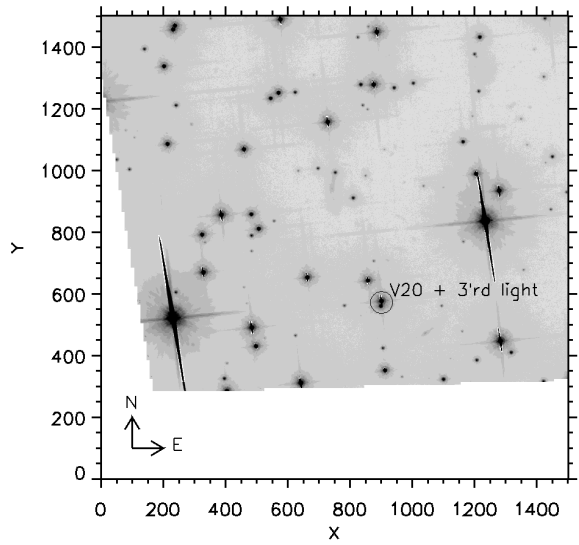}
\caption[]{\label{fig:V20fc}
Finding chart for V20. Plate scale is 0.05\arcsec per pixel. Image from ACS on {\it HST}}
\end{figure*}

\begin{figure*}
\centering
\epsfxsize=80mm
\epsfbox{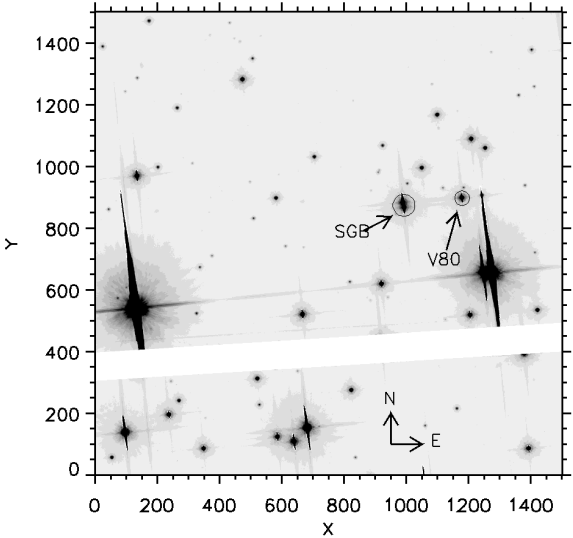}
\caption[]{\label{fig:V80fc}
Finding chart for V80 and SG. Plate scale is 0.05\arcsec per pixel. Image from ACS on {\it HST}}
\end{figure*}

\end{document}